\documentclass[a4paper,fleqn,usenatbib]{mnras}

\usepackage[T1]{fontenc}
\usepackage{ae,aecompl}

\usepackage{graphicx}	
\usepackage{amsmath}	
\usepackage{amssymb}	
\usepackage{multicol}        
\usepackage{bm}		
\usepackage{pdflscape}	
\usepackage{hyperref}

\usepackage{float}
\usepackage{color,soul}
\usepackage{xcolor}

\newcommand{\lya}{Lyman-$\alpha$}
\newcommand{\angstrom}{\mbox{\normalfont\AA}}

\title[BAO at high z]{On the possibility of Baryon Acoustic Oscillation measurements at redshift $z>7.6$ with the Roman Space Telescope}

\author[S. Satpathy et al.]{
Siddharth Satpathy,$^{1,2}$\thanks{E-mail: siddharthsatpathy.ss@gmail.com}
Zhaozhou An,$^{1,2}$
Rupert A. C. Croft,$^{1,2,3,4}$
\newauthor
Tiziana Di Matteo,$^{1,2,3,4}$
Ananth Tenneti,$^{1,2}$
Yu Feng,$^{5}$
Katrin Heitmann$^{6}$
\newauthor
\& Graziano Rossi$^{7}$
\\
$^{1}$Department of Physics, Carnegie Mellon University, 5000 Forbes Ave., Pittsburgh, PA 15213, USA\\
$^{2}$The McWilliams Center for Cosmology, Carnegie Mellon University, 5000 Forbes Ave., Pittsburgh, PA 15213, USA\\
$^{3}$ School of Physics, The University of Melbourne, VIC 3010, Australia\\
$^{4}$ ARC Centre of Excellence for All Sky Astrophysics in 3 Dimensions (ASTRO 3D), Australia\\
$^{5}$Berkeley Center for Cosmological Physics, Department of Physics, University of California Berkeley, Berkeley, CA 94720, USA\\
$^{6}$HEP and MCS Divisions, Argonne National Laboratory, Lemont, IL 60439, USA\\
$^{7}$Department of Physics and Astronomy, Sejong University, Seoul, 143-747, Korea
}

\date{Accepted XXX. Received YYY; in original form ZZZ}

\pubyear{2019}

\DeclareUnicodeCharacter{2212}{-}
\begin{document}
\label{firstpage}
\pagerange{\pageref{firstpage}--\pageref{lastpage}}
\maketitle

\begin{abstract}
The Nancy Grace Roman Space Telescope (RST), with its field of view and high sensitivity will make surveys of cosmological large-scale structure possible at high redshifts. We investigate the possibility of detecting Baryon Acoustic Oscillations (BAO) at redshifts $z>7.6$ for use as a standard ruler. We use data from the hydrodynamic simulation \textsc{BlueTides}  in conjunction with the gigaparsec-scale Outer Rim simulation and a model for patchy reionization to create mock RST High Latitude Survey grism data for \lya\ emission line selected galaxies at redshifts $z=7.4$ to $z=10$, covering 2280 square degrees.  We measure the monopoles of galaxies in the mock catalogues and fit the BAO features. We find that for a line flux of $L = 7\times 10^{-17} \ {\rm erg/s/cm}^{2}$, the  $5 \sigma$ detection limit for the current design, the BAO feature is partially detectable (measured in three out of four survey quadrants analysed independently). The  resulting root mean square error on the angular diameter distance to $z=7.7$ is 7.9$\%$.  If we improve the detection sensitivity by a factor of two  (i.e. $L = 3.5\times 10^{-17} \ {\rm erg/s/cm}^{2}$), the distance error reduces to $1.4\%$. We caution that many more factors are yet to be modelled, including dust obscuration, the damping wing due to the intergalactic medium, and low redshift interlopers. If these issues do not strongly affect the results, or different observational techniques (such as use of multiple lines) can mitigate them, RST or similar instruments may be able to constrain the angular diameter distance to the high redshift Universe.
\end{abstract}



\section{Introduction}  \label{sec: intro}
Baryon Acoustic Oscillations (BAO) have become very useful as a standard ruler to constrain cosmological models (see e.g., the review by \citet{Weinberg2013}). Due to the distinctive nature of their signature, the BAO scale constitutes one of the most effective observables in cosmology. The BAO features can be detected as peaks in two point correlation functions or wiggles in power spectra. These signatures of BAO can be used to measure the Hubble constant and the angular diameter distance as function of redshift. As such, it is used as a robust and unique probe to investigate the nature of dark energy.

The sound horizon at the time of recombination leaves its imprint on both the cosmic microwave background (CMB) and three dimensional distribution of matter. Using simulations, in this paper we explore whether it may be possible to measure the BAO scale at redshifts $z>7.6$ using galaxies detected by the RST satellite (formerly the Wide Field Infrared Survey Telescope, WFIRST\footnote{ \href{https://wfirst.gsfc.nasa.gov/}{https://wfirst.gsfc.nasa.gov/} }). These measurements would represent the first BAO detection between $z=2.3$ and $z=1100$.

The first observations made of BAO \citep[][]{Eisenstein2005, Cole2005}) at low redshifts ($z\sim0-0.4$) used luminous red galaxies (LRGs) as tracers of structure, and these have been supplemented by even larger samples at higher redshifts (e.g., \citet{Zhai2017}, \citet{Bautista2018}), up to $z=0.72$. The density field at redshifts $z \sim 2-3.5$ is accessible using the \lya\ forest of absorption in quasar spectra, and its measured clustering has led to the determination of the BAO scale at $z \sim 2.3 - 2.4$ \citep[][]{Busca2013, Slosar2013, duMas2017, deSainte2019}. The gap between low redshift galaxy determinations and the forest was recently filled by a determination using quasar clustering at $z=1.52$ \citep[][]{Ata2018}. Emission line galaxies at $z=0.85$ \citep[][]{Raichoor2017} and $z=0.9$ \citep[][]{Comparat2016} from the Extended Baryon Oscillation Spectroscopic Survey (eBOSS) will also soon be adding constraints in this crucial redshift range, around the time that the Universe started to accelerate. Radio observations of 21cm emission will increase the volume covered \citep[][]{Vanderlinde2014}. At higher redshifts, galaxies have been discovered as far away as $z=11.1$ \citep[][]{Oesch2016}, in small Hubble Space Telescope (HST) fields at present. The RST satellite$^{1}$ will have a much wider field of view than HST, and the planned RST grism survey holds the promise of measuring the large-scale structure of the Universe at early times with galaxy survey techniques which have been hitherto been confined to low redshifts.

Among the reasons for looking at the BAO scale in a different redshift regime are the possibility of early dark energy \citep[][]{Hill2018}, or models where dark matter decays into radiation and affects the expansion rate \citep[][]{Wang2014}. There is also current tension between locally determined $H_{0}$ measurements \citep[][]{Riess2018} and those determined from the CMB in the context of cold dark matter (CDM) \citep[][]{Planck2018}. Using multiple redshifts to determine the consistency (or not) of BAO measurements should be useful in finding a solution to this controversy.

Cosmological surveys designed to measure the BAO feature must cover large volumes, as the comoving length scale involved is 150 megaparsecs, and cosmic variance must be suppressed. The number density of sampling objects must be large enough to overcome shot noise. When there is domination by shot-noise, one can still obtain BAO measurements and the error will just scale as $1 / \sqrt(N)$, where $N$ is the number of objects. For example, eBOSS quasars are shot-noise dominated \citep[][]{Zarrouk2018}. The shot-noise dominated regime carries the implication that one can return to the same volume and sample it further to obtain more precise results. Conversely, when one is cosmic variance limited, one gains more from sampling a new volume instead of sampling the same volume.

The current number density of quasars at $z\sim3$ in the largest survey \citep[final SDSS-III BOSS DR09;][]{Eftekharzadeh2015} is $1.057 \pm 0.007$ ($h^{-1}$Mpc)$^3$, which is too low. Galaxies at similar redshifts and higher exist in sufficient numbers, but their relatively low luminosity has made gathering samples that cover sufficient volume impossible with current technology. RST and Euclid will enable the Universe at $z=6$ to be studied with a more representative sample of galaxies than have been possible at present (small current fields of view lead to rarer brighter galaxies being missed). So far, galaxies have been found up to redshifts $z=12$ with dropout techniques \citep[][]{Ellis2013}. The largest samples with photometric redshifts at $z \geq 6$ are \citet{Chaikin2018}, \citet{Yan2011}, \citet{Bouwens2011}, \citet{Laporte2017}, \citet{Eyles2005} and \citet{Mobasher2005}  which contain 72, 20, 1, 1, 2 and 1 galaxies respectively. In a few cases spectral confirmation has been obtained by identifiying the \lya\ line (refs for $z>7.5$ galaxies  \citep{Finkelstein2013, Song2016}. Between $z=6$ and $7$, the Candels survey \citep[][]{Pentericci2018} has found 260 spectroscopically confirmed galaxies. Narrow band imaging surveys searching directly for the \lya\ line offer a route to large samples in principle.  An early example of such a search includes \citet{Tilvi2010} at $z=7.7$ which was not able to confirm any objects. More recently, the LAGER survey \citep[][]{Hu2017} has followed up candidates with further spectroscopy and found a 66\% success rate at $z\sim7.0$.

There are  number of obstacles to the finding and use of $z>7.6$ galaxies as a BAO probe. The most important is the actual visibility of lines that would enable spectroscopic identification. The \lya\ line is subject to multiple scatterings and absorption by dust within galaxies. It is also absorbed by neutral hydrogen content of the intergalactic medium (IGM), and \lya\ emitting galaxies at the highest redshifts have proved difficult to find. One the other hand, at $z\sim 6.5$, \citet{Bagley2017} have found that  \lya\ emitters are detectable and may lie mainly in ionized bubbles, sufficiently large for the \lya\ photons to redshift out of resonance before encountering the neutral IGM gas. At lower redshifts (e.g., $z\sim2-3$), \lya\ emission appears to have an escape fraction of $\sim 5-10$\% \citep[][]{Gronwall_2007a, Francis2012} but at higher redshifts $z>6$ it appears to climb to near 100\% \citep[][]{Cassata2011, Sobacchi2015}.

Interlopers are another important issue. For example there will be galaxies at redshifts $1.1 < z< 1.9$ for which Hydrogen Balmer lines will fall in the spectral window for \lya\ at redshifts $\sim8$. Intermediate redshift ($1.7 < z< 2.8$) galaxies will also contribute [OIII] 5007 emission. How well RST grism spectra could be used to identify galaxies at the highest redshifts and remove interlopers is an open question.

In this paper we carry out the simplest analysis possible, using a combination of a large darkmatter simulation \citep[Outer Rim;][]{Li2019} and a hydrodynamic simulation \citep[\textsc{BlueTides};][]{Wilkins2016a, Wilkins2016b} to see whether BAO could be detected at redshifts $z \sim 8$ with RST's spectral coverage. We make many simplifying assumptions, each of which are likely to strongly affect the outcome, but can be tested with future work. We assume that all \lya\ emission can be seen. In other words, we suppose that the \lya\ emitting galaxies lie in ionized bubbles \citep[][]{Kakiichi2016, Yajima2018}. We ignore galaxy-scale extinction of \lya\ emission. That is, we assume that the raw star formation in our hydrodynamic simulation of galaxy formation, \textsc{BlueTides} \citep[][]{Wilkins2016a, Wilkins2016b} can be directly translated into \lya\ luminosity. We also do not directly address interlopers and do not do a complete analysis of spectral overlap in the grism spectra. We restrict ourselves to seeing whether mock observations of a RST grism survey under these conditions can detect the BAO feature. Our analysis therefore acts as a first step - if we find that it is possible, then these restrictions should be relaxed and tested to see whether it is worth pursuing the idea further.  

The plan for the paper is as follows. In Section~\ref{sec: baosimulationdata} we describe the different cosmological datasets we use. In Section~\ref{sec:BAOMethodology} we show how these were used to construct mock catalogues for the RST high latitude survey. We describe details results obtained from mock catalogues in Section~\ref{sec:bao_results}. We present comprehensive discussion and analysis of our results in Section~\ref{sec: baodiscussion}.

\begin{table}
    \centering
    \caption{Simulation cosmological parameters based on assumed $\Lambda$CDM cosmological model.}
    \label{tab:cos_bluetides}
    \begin{tabular}{llll}
        \hline
        Name & \textsc{BlueTides} & Outer Rim \\
        \hline
        $\Omega_{\Lambda}$ & 0.7186 & 0.7352 \\
        $\Omega_{\text{Matter}}$ & 0.2814 & 0.2648\\
        $\Omega_{\text{Baryon}}$ & 0.0464 & 0.04479\\
        h &0.697 &0.71 &\\
        $\sigma_{8}$ & 0.820 & 0.8\\
        $n_s$ & 0.971 &0.963\\
        \hline
    \end{tabular}
\end{table}



\section{Datasets} \label{sec: baosimulationdata}

\subsection{The \textsc{BlueTides} simulation} \label{sec: bluetidesdata}

\textsc{BlueTides} is a hydrodynamic simulation which was designed to cover the high redshift Universe \citep[][]{Feng2015, Feng2016}. The model uses the Smoothed Particle Hydrodynamics code \textsc{MP-Gadget} to evolve a cubical volume with $2 \times 7040^3$ particles. These particles are used to model dark matter, gas, stars and supermassive black holes. This simulation evolved a $(400/h \approx 577 )^3$ cMpc$^3$ cube to redshift $z=7.3$. The resolution, volume and number of galaxy halos in \textsc{BlueTides} are well matched to depth and area of current observational surveys targeting the high-redshift Universe. Additional details of the \textsc{BlueTides} simulation can be found in \citep[][]{Feng2015, Feng2016, Wilkins2016a, Wilkins2016b, Wilkins2017}.

A Friends-of-Friends (FoF) algorithm was used to identify galaxy halos. At the high redshifts relevant here these halos are sufficiently accurate representations of the galaxies that would be found in the simulations if observational determinations were used. This was seen  directly in \citet{Feng2016} by comparing FoF halos with galaxy halos extracted from \textsc{BlueTides} mock images and identified with the Source Extractor algorithm \citep[][]{Bertin1996}.

\citet{Feng2016} used results from the \textsc{BlueTides} simulation to find good correspondence between the predicted galaxy luminosity function and observations in the observable range $-18 \leq M_{\rm UV} \leq -22.5$  with some dust extinction required to match the abundance of brighter objects. Additionally, \citet{Feng2016} also find good agreement between the star formation rate density in \textsc{BlueTides} and current observations in the redshift range $8 \leq z \leq 10$. 

\cite{Wilkins2017} use galaxy halos from \textsc{BlueTides} simulation to show that intrinsic and observed luminosity functions at $ z \in \lbrace 8, 9, 10  \rbrace $ exhibit the fast expected build up of the galaxy population at high-redshifts. The work presented in \cite{Wilkins2017} reveals that the intrinsic luminosity function is broadly similar to the observed luminosity function at faint luminosities ($ M > -20$). At brighter luminosities, the luminosity function is shown to have a stronger steepening which indicates the increasing strength of dust attenuation. In their work, \cite{Wilkins2017} also demonstrate that a Schechter function provides a good overall fit to the shape of a  the dust attenuated far-UV luminosity function for $ z \in \lbrace 8, 9, 10  \rbrace $.

In \cite{waters2016}, it was shown that the number density of galaxies that will be observable by the RST High Latitude Survey should be high enough that BAO measurement at redshift $z\sim8$ could be possible. The \textsc{BlueTides} volume being too small to make direct mock HLS observations and measurements of the BAO, it was necessary to follow up with different techniques, and this is the role of the current paper.

The \textsc{BlueTides} simulation is used in this paper in two roles:  First, to make some example RST grism fields to check the degree of spectral overlap between high redshift galaxies and others at similar redshifts. Second, we use the \textsc{BlueTides} galaxies to populate the dark matter halos in a much larger dark matter only simulation. For the dark matter only simultion, we use the Outer Rim catalogue. Details of the Outer Rim simulation are described in ~\ref{sec: outerrimdata}.

\subsection{The Outer Rim simulation} \label{sec: outerrimdata}

\begin{figure}
    \includegraphics[width=\columnwidth]{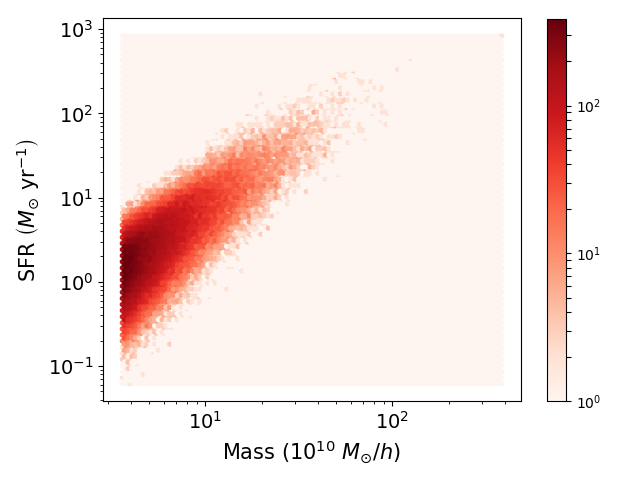}
    \caption{The variation of star formation rate with mass for \textsc{BlueTides} galaxies at $z=8$.}
    \label{fig:mass_sfr}
\end{figure}

\begin{figure}
    \includegraphics[width=\columnwidth]{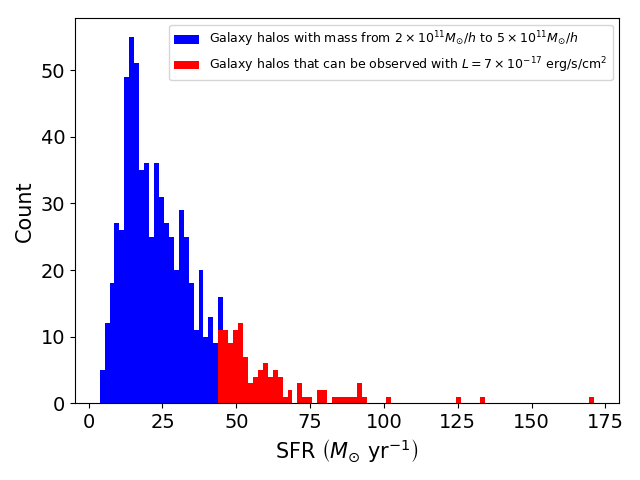}
    \caption{A histogram of star formation rates for galaxies with masses from $2\times 10^{11}M_{\sun}/h$ to $5\times 10^{11}M_{\sun}/h$ in redshift 8. The red bins represent galaxies that can be observed with line detection sensitivity, $L = 7\times 10^{-17} \ {\rm erg/s/cm}^{2}$.}
    \label{fig:sfr_dis_mass20to50} 
\end{figure}
    
The Outer Rim simulation \citep[][]{Heitmann2019, Li2019} is a large-scale dark matter only simulation which we use to measure large-scale clustering and BAO. It was run using  the Hardware/Hybrid Accelerated Cosmology Code \citep[HACC;][]{Habib2013, Habib2016}, on Mira, a BG/Q system at the Argonne Leadership Computing Facility. The cosmology used is a $\Lambda$CDM model close to the best-fit model from WMAP-7 \citep[][]{Komatsu2011}, with parameters given in Table~\ref{tab:cos_bluetides}. The comoving box size of the simulation is  $3000 \ h^{-1}$Mpc, yielding a volume 400 times that of \textsc{BlueTides}. 1.07 trillion particles were evolved, leading to a particle mass of $m_{p} = 1.85 \times 10^{9} \ h^{-1} M_{\sun}$. The model was run to redshift $z=0$, but in this paper we use only outputs at $z=7$ and above.

Halos are identified with a Friends-of-Friends \citep[][]{Davis1985} halo finder with a linking length of $b = 0.168$.

\subsection{Galaxy Lyman-$\alpha$ flux} \label{sec: gal_lya}

\begin{figure}
    \includegraphics[width=\columnwidth]{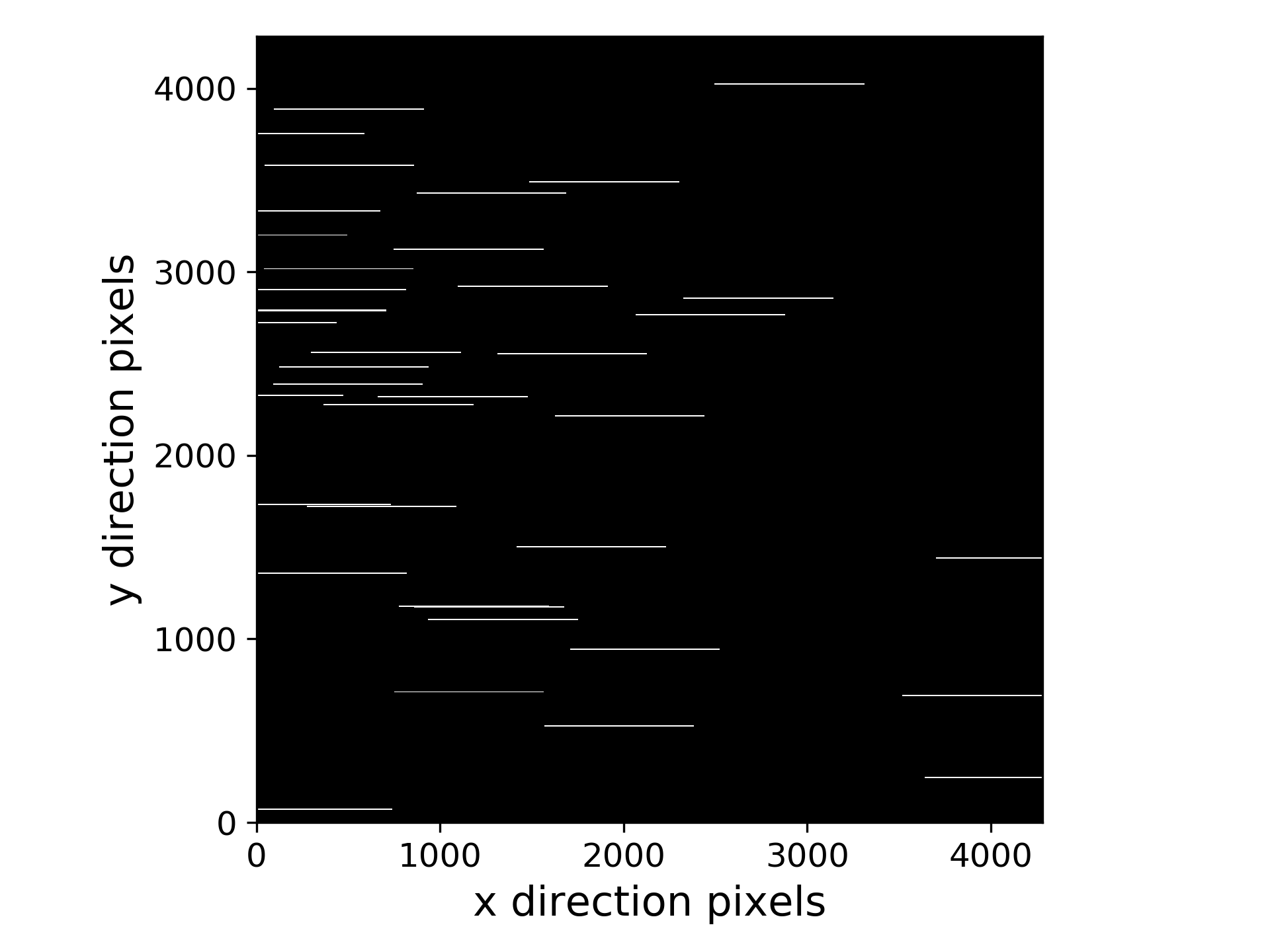}
    \caption{A simulation of a grism image with galaxies at redshift $z=8$ from the \textsc{BlueTides} simulation. Each horizontal line denotes the spectrum of a galaxy. The brightness is proportional to the flux in the pixels.}
    \label{fig:grism_simu}
\end{figure}

At the redshifts $z \sim 8$ which concern us here, we will simulate the \lya\ line luminosity of galaxies, which will lie within the RST grism spectral range (see section~\ref{sec: mock_grism} below). An example of \lya\ line detection from a galaxy at these redshifts is given in the work of \citet{Lehnert2010}. The redshift of the galaxy UDFy-381355395 is $z=8.6$, and \citet{Lehnert2010} measured a \lya\ line flux of  $(6.1\pm1.0)\times 10^{−18}$ erg/s/cm$^{2}$, corresponding to a luminosity of $5.5\pm1.0\pm1.8\times 10^{42}$ erg/s.

In \textsc{BlueTides}, the star formation rate (SFR) for each galaxy is available to us, and we translate this quantity directly into \lya\ luminosity using the following scaling equation \citep[][]{Cassata2011}.

\begin{equation}
\label{eqn:sfr_lyalpha}
L_{ {\rm Ly } \alpha} ( {\rm erg \ s^{-1}} ) =1.1\times 10^{42} \times {\rm SFR} ( M_{\sun} \ {\rm yr}^{-1} )
\end{equation}

In equation~\ref{eqn:sfr_lyalpha},  $L_{ {\rm Ly } \alpha}$ denotes the \lya\ luminosity. This conversion factor is based on a stellar population with a Salpeter initial mass function (IMF) and is accurate to within a factor of a few for a range of population ages, high mass cutoffs of stars, and metallicities \citep[][]{Leitherer1999}. Most relevant in our case is that there is no correction for effects like dust,  escape fraction, and intergalactic or circumgalactic absorption. This means that using the raw star formation rate measurements from the simulation represents a best case for the detectability of galaxies and this should be borne in mind when interpreting the results. We will return to  this in Section~\ref{sec: baodiscussion}. Due to the paucity of observational data at these redshifts, the magnitude of these obscuring effects is not well known, but they are likely to be significant. In the example of the galaxy mentioned in \citet{Lehnert2010}, the continuum SFR was estimated to be $2-4 \ M_{\sun} \ {\rm yr}^{-1}$. At the same time, \citet{Lehnert2010} evaluated the SFR from the \lya\ line to be $0.3-2.1 \ M_{\sun} \ {\rm yr}^{-1}$, which leads to \lya\ overall escape fraction from $8\%-100\%$.

\begin{figure}
    \includegraphics[width=\columnwidth]{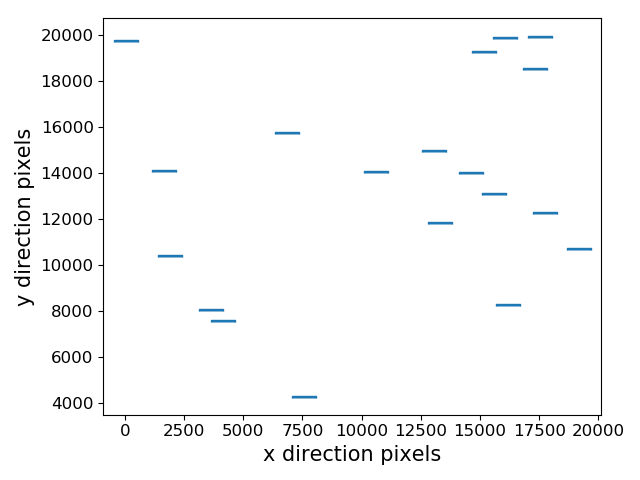}
    \caption{A grism image for a model for $L = 7\times 10^{-17} \  {\rm erg/s/cm}^{2}$. Each blue block represents a grism spectrum for a single galaxy. We use this plot to estimate overlap.}
    \label{fig:grism_image}
\end{figure}

The \textsc{BlueTides} simulation volume is not sufficiently large to make mock RST survey catalogues that can be used to measure large-scale structure. We use the halo catalogues from the Outer Rim simulation to do this, painting SFRs  from \textsc{BlueTides} galaxies
onto correspondingly massive Outer Rim halos. Fig.~\ref{fig:mass_sfr} shows the relationship between mass and SFR for \textsc{BlueTides} at redshift $z=8$, We can see that there is an approximately linear relationship between $\log (M)$ and $\log ({\rm SFR})$. Our SFR assignment scheme is as follows: for each Outer Rim halo, we randomly pick a \textsc{BlueTides} halo with total mass within $5 \times 10^{10}$ $h^{-1}M_{\sun}$, and assign the SFR for that halo to the Outer Rim object. In this way, we account for some aspects of the scatter between SFR and halo mass, although not any spatially correlated structure. For large mass galaxies we increase the range to make sure there are enough sample halos in \textsc{BlueTides}.

In one of the RST design documents, the $7 \sigma$ line flux sensitivity for the High Latitude Survey is $10^{-16} \ {\rm erg/s/cm}^{2}$. \footnote{  \href{https://wfirst.gsfc.nasa.gov/science/sdt_public/WFIRST-AFTA_SDT_Report_150310_Final.pdf}{https://wfirst.gsfc.nasa.gov/science/sdt$\_$public/WFIRST-AFTA$\_$SDT$\_$Report$\_$150310$\_$Final.pdf}  } We discuss RST and the flux limits on our mock surveys further in Section~\ref{sec: RST_grism}. 

Fig.~\ref{fig:sfr_dis_mass20to50} shows the SFR distribution of \textsc{BlueTides} galaxies at $z=8$ with mass from $2 \times 10^{11} \ h^{-1} M_{\sun}$ to $5 \times 10^{11} \ h^{-1} M_{\sun}$. We can see that when the line sensitivity is $7\times 10^{-17} \ {\rm erg/s/cm}^{2}$, then the minimum detectable luminosity is $5\times 10^{43} \ {\rm erg/s}$ at $z=8$, which corresponds to $40 \ M_{\sun} \ {\rm yr}^{-1}$. Only the most highly star-forming galaxies are therefore likely to be included in any BAO survey.

\section{Methodology} \label{sec:BAOMethodology}

\begin{figure}
    \includegraphics[width=\columnwidth]{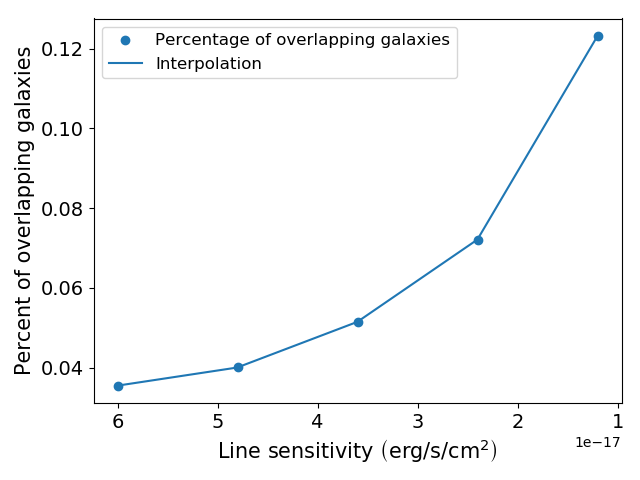}
    \caption{The percentage of overlapping galaxies as a function of line sensitivity. We notice a monotonically increasing trend in the variation of the percent of overlapping galaxies with line sensitivity. The solid line denotes a one dimensional interpolation between data points.}
    \label{fig:overlap_L}
\end{figure}

\subsection{Mock grism images} \label{sec: mock_grism}

Working with grism spectroscopy is complex \citep[][]{McCarthy1999, Fossati2017, Colbert2018}, with relatively low resolution possible, mixing of spatial and spectral structure in background and stray light, and overlap of sources. We leave full simulation of mock RST grism surveys (including foregrounds and interlopers) to other work \citep[][]{Gehrels2015, Colbert2018, Malhotra2019}. We will however make some extremely simple grism images using \textsc{BlueTides} simulation data, in order to assess the level of spectral overlap from galaxies at similar redshifts.

\subsubsection{The RST High Latitude Grism Survey} \label{sec: RST_grism}

The RST High Latitude Survey (HLS) is planned to have a nominal sky coverage of 2400 deg$^2$, which will greatly increase the number of galaxies available at redshifts $z \geq 8$. This survey will be conducted by RST's Wide Field Instrument (WFI) and is expected to be a dual imaging and grism spectroscopy  survey. The imaging survey will obtain direct imaging through four filters (Y106, J129, H158, and F184) and grism spectroscopy will be done at four roll angles. The filters Y106, J129, H158, and F184 are expected to enable imaging in wavelength ranges $0.927-1.192 \ \mu {\rm m}$, $1.131-1.454 \ \mu {\rm m}$, $1.380-1.774 \ \mu {\rm m}$ and $1.683-2.000  \ \mu {\rm m}$ respectively. The imaging survey will therefore cover the wavelength range ($0.927-2 \ \mu {\rm m}$). The
grism coverage is the most relevant for our work, and the planned accessible range of
wavelengths specified in the WFIRST-AFTA SDT report 2015 (\citep{spergel2015}) is $1-1.93 \ \mu {\rm m}$. With these specifications, \lya\ emission at $ 7.22 < z < z=14.88$ will be visible.

\begin{figure}
    \includegraphics[width=\columnwidth]{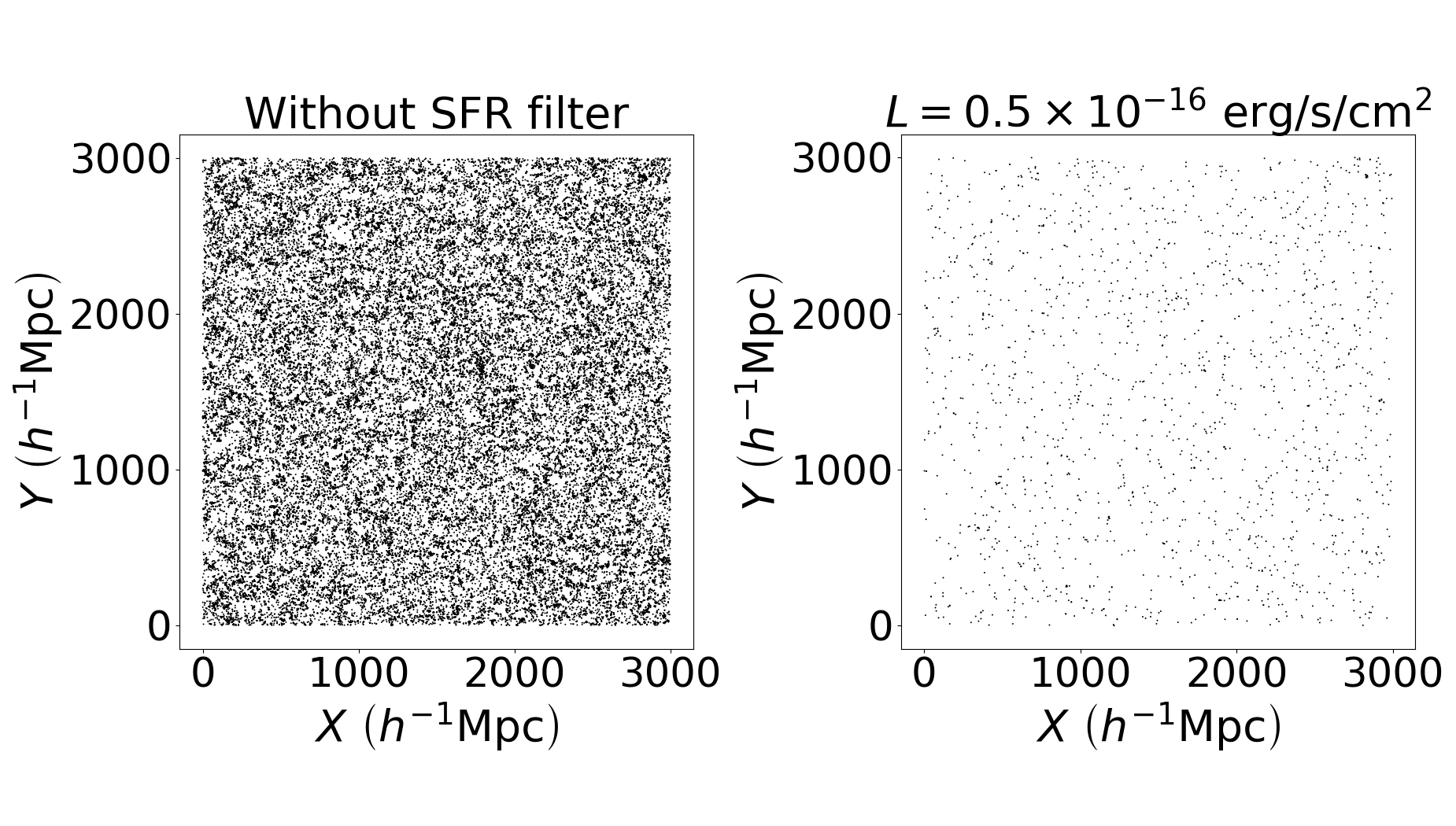}
    \caption{The figure on the left is cross section of original data. The figure on the right is the cross section for same position in simulated data with $L = 6\times 10^{-17} \ {\rm erg/s/cm}^{2}$.}
    \label{fig:cross_section}
\end{figure}

\subsubsection{Image construction} \label{sec: image_reconstruction}

We take the $z=8$ snapshot of \textsc{BlueTides} and project the positions of galaxies into two dimensions using Euclidean geometry. The simulation box comoving side length of $400\ h^{-1}$Mpc corresponds to the comoving radial distance between $z=7.5$ and $z=9.5$. As the number density of galaxies decreases rapidly with redshift, our estimate of the overlap between galaxies in our crude image simulation is likely to be conservative. At each galaxy position we plot a dispersed galaxy spectrum using the relationship between angular position and wavelength appropriate for RST ($1 - 2 \ \mu {\rm m}$) .
We include the effect of the redshift distortions on mock grism image by shifting the wavelengths of the spectrum of each galaxy according to the peculiar velocity along the line of sight.
The galaxy spectrum data corresponds to models described in Fig. 2d of \citet{Wilkins2013b}. Fig. 2d of \citet{Wilkins2013b} illustrates the spectral energy distribution of a stellar population which has formed stars at a constant rate for 100 Myr for $z=8.0$. For our present purposes identical spectra for each galaxy are sufficient and we do this for ease of computation. In future work with more realistic simulations, the spectra for each individual galaxy can be determined using population synthesis models which are available and could be used \citep[][]{Wilkins2019}.

The angular extent of the spectra perpendicular to the dispersion direction is determined by the approximate angular size of each individual galaxy. For the angular size part, we refer to \citet{Feng2016} to get ${\rm M_{UV}}$ from SFR. After this we use relationship of galaxy half-light radii vs ${\rm M_{UV}}$ shown in Figure 4 of \citet{Feng2015} to get galaxy half life radius from ${\rm M_{UV}}$. Using this information, we find that galaxies of mass $10^{11} \ h^{-1} M_{\sun}$ have an angular size of 0.11 arcsec in the grism image.

Fig.~\ref{fig:grism_simu} shows one of the simulated grism images. The brightness shown in the picture is proportional to the flux in the pixels.

\subsubsection{Spectral overlap} \label{sec: spectral_overlap}

It is common that there will be overlap between galaxies in the grism image because some galaxies are close to each other and the direction is close to the direction between two galaxies. This will make the process of extracting galaxy spectra harder and lead to a decrease in the number of observed galaxies with reliable position information. It is therefore necessary to consider the overlap in the grism image. Usually we can mitigate this effect by taking grism images with different dispersion direction. RST will take grism images at four angles: $0^{\circ}$, $15^{\circ}$, $180^{\circ}$ and $195^{\circ}$. To estimate the upper bound of overlap, we will only consider a grism image with dispersion in one direction.  Fig.~\ref{fig:grism_image} shows the simplified model that we use to estimate overlap. We use rectangular blocks that have the same length as grism spectrum to replace the galaxies, and find the blocks that overlap with each other and mark the galaxies as overlapping galaxies. We calculate the percentage of overlapping galaxies. Fig.~\ref{fig:overlap_L} shows the relationship between the percentage of overlapping galaxies and line sensitivity. We can see the percentage increase when line sensitivity decreases, with the percentage around $7\%$ at $L = 3.5\times 10^{-17} \ {\rm erg/s/cm}^{2}$ (we will see later that this is around the level that accurate BAO fitting is possible).  We calculate the upper bound of overlap by neglecting multiple dispersion directions and regarding partial overlap as full overlap, so the real overlap should be less than this. 

We note of course that here we are only treating overlap from galaxies in the same high redshift range that we are interested in. These will be greatly outnumbered by  a much higher density of foreground galaxies from below $z=7.4$ in an actual survey. We leave the problem of overlap from these galaxies and also non-overlapping interlopers (galaxies with other spectral lines in the same wavelength range as \lya\ at high redshifts) to future work.

\subsection{Mock survey data} \label{sec: mock_survey}

\begin{figure}
    \includegraphics[width=\columnwidth]{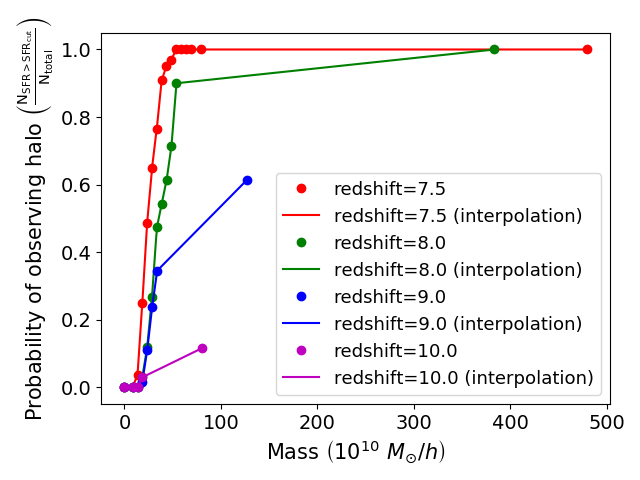}
    \caption{The relationship between the probability of a galaxy being observed and its mass. The solid lines are one dimensional interpolations between data points.}
    \label{fig:p_mass}
\end{figure}

The data we use from the Outer Rim simulation covers snapshots from $\text{z}=7.4$ to $\text{z}=10$. Our mock RST high-latitude survey (HLS) data, will cover four $3000 \times 3000 \times 700 \ h^{-3}$Mpc$^{3}$ cuboids drawn from the $3000^{3} \ h^{-3}$Mpc$^{3}$ simulation volume. This corresponds to a sky area of $4 \times 570=2280$ sq. deg, approximately equal to the total sky coverage of the RST HLS. The redshift coverage in the line of sight dimension is $\text{z}=7.4$ to $\text{z}=10$, with the survey volume covering $4\times6.3=25.2$ comoving (Gpc/$h)^{3}$. We use the flat sky approximation for simplicity, and the conversion from distance to angle above refers to the edge of the volume, at $z=10$. Because the whole of the Outer Rim simulation is necessary to produce a single mock RST HLS, we will analyse the four separate cuboids drawn from Outer Rim separately, as quarter-mocks. The scatter between the results from the different quarter-mocks will give us a crude estimate of cosmic variance.

In order to include redshift evolution in the mock surveys, we choose a cuboid with fixed position in different snapshots, and calculate the comoving radial distance of each redshift.  We then extract the corresponding slice in the cuboid from different snapshots according to the relevant comoving radial distance. Stacking all the slices in  order of redshift yields the mock survey. The redshift evolution is discrete rather than a true lightcone - we use redshifts at z=7.4, 7.7, 7.9, 8.4, 8.6, 8.8, 8.0, 9.2, 9.4, 9.6, 9.8 and 10.0. We include the effect of redshift distortions in the mock surveys using
the peculiar velocities of the simulated galaxies.

As the Outer Rim  simulation covers a $3000\times 3000 \times 3000 \ h^{-3}$Mpc$^{3}$ cube, we are able to create four mock cuboids from the Outer Rim data, using different parts of the volume. Each of these covers $\sim 1/4$ of the RST HLS area, so that there are therefore four approximately independent quarter-mock surveys to use for baryon acoustic peak fitting.

\begin{figure}
    \includegraphics[width=\columnwidth]{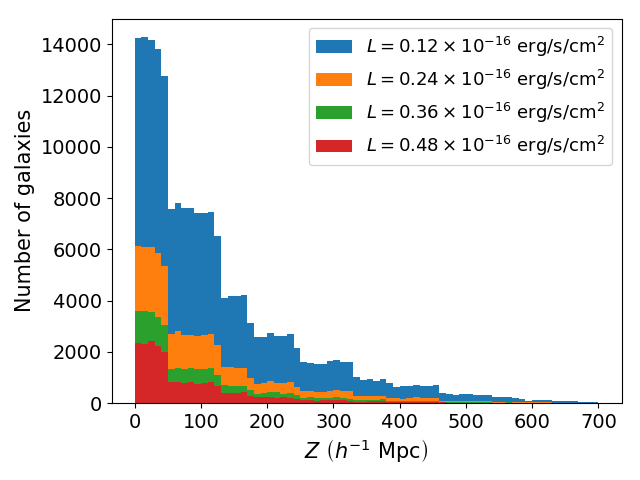}
    \caption{Here we show the distribution of galaxies along the z-axis for different line sensitivities.}
    \label{fig:number_z_L}
\end{figure}

As stated above, we assign \textsc{BlueTides} galaxy luminosities to Outer Rim halos according to their mass. We make several sets of mock surveys corresponding to different galaxy detection thresholds (different line sensitivities): \\
\\

\begin{table*}
	\centering
	\caption{This table shows the numbers of halos present in, and the mean redshifts of the three quarter-mock catalogues (L7, L3.5, L1).}
	 \label{tab:num_halos_mean_readshift}
	\begin{tabular}{lrrrr}
		\hline
		\hline
    		\textbf{Line detection} & \textbf{Mock 1} & \textbf{Mock 2}  & \textbf{Mock 3}  & \textbf{Mock 4}  \\
    		\textbf{sensitivity} & & & &  \\ 
    		\hline
    		\hline
    		\textbf{Number of halos (before patchy reionization)} & & & & \\
    		\hline    		
    		$7$ erg/s/cm$^2$ & 15,484 & 15,344 & 15,280 & 15,144 \\        
    		$3.5$ erg/s/cm$^2$ & 37,091 & 37,091 & 37,154 & 36,896 \\        
    		$1$ erg/s/cm$^2$ & 194,684 & 194,522 & 195,157 & 194,107 \\
    		\hline
    		\textbf{Number of halos (with patchy reionization)} & & & & \\
    		\hline
    		$7$ erg/s/cm$^2$ & 13,999 & 14,040 & 13,989 & 13,910 \\        
    		$3.5$ erg/s/cm$^2$ & 32,124 & 32,238 & 32,465 & 32,232 \\        
    		$1$ erg/s/cm$^2$ & 147,463 & 148,150 & 148,994 & 147,529 \\ 
    		\hline
    		\textbf{Mean redshifts (before patchy reionization)} & & & & \\
    		\hline
    		$7$ erg/s/cm$^2$ & 7.6474 & 7.6498 & 7.6530 & 7.6519 \\        
    		$3.5$ erg/s/cm$^2$ & 7.6895 & 7.6912 & 7.6962 & 7.6953 \\        
    		$1$ erg/s/cm$^2$ & 7.7852 & 7.7863 & 7.7884 & 7.7867 \\ 
    		\hline
    		\textbf{Mean redshifts (with patchy reionization)} & & & & \\
    		\hline
    		$7$ erg/s/cm$^2$ & 7.6645 & 7.6650 & 7.6657 & 7.6711 \\        
    		$3.5$ erg/s/cm$^2$ & 7.7156 & 7.7168 & 7.7197 & 7.7228 \\        
    		$1$ erg/s/cm$^2$ & 7.8276 & 7.8280 & 7.8294 & 7.8297 \\    		    		   		    		   		    		
    		\hline
    		\hline    		
	 \end{tabular}
\end{table*}

(a) L7 quarter-mocks. The 2015 WFIRST technical report \citep[][]{Casertano2015} estimates that with current design parameters, the RST HLS grism survey will have a $7 \sigma$ H$\alpha$ line detection sensitivity of $6-12 \times 10^{-17} \ {\rm erg/s/cm}^{2}$ for sources with effective radius $<0.3$ arcsecs. This is for H$\alpha$ spectral line ($\lambda_{\rm rest}=6563 \ \angstrom $) in the redshift range $z=1.05$ to $z=1.88$. In \cite{spergel2015}, a value of $1 \times 10^{-17} \ {\rm erg/s/cm}^{2}$ at $7 \sigma$ is quoted, which is in the same range. In our analysis, we use this value, but assume a 5 $\sigma$ detection limit, which results in a lower limit to the detectable line flux of $7 \times 10^{-17} \ {\rm erg/s/cm}^{2}$. As the RST survey strategies are as yet unfinalized, we assume that there could be some flexibility, both in the selection of the fiducial sensitivity, and also possibilities
of deeper integrations over part of the survey area.

We assume that this sensitivity can be applied to the \lya\ line over the same observed range of wavelengths. The line sensitivity for resolved sources is approximately a factor of two worse, but we expect the galaxies to be unresolved. For example, \citet{Feng2015} find half light radii of $0.5\sim1$ kpc for the most massive galaxies in \textsc{BlueTides} at redshift $z=8$, which corresponds to $0.1-0.2$ arcsec. For our first mocks, those closest to the RST design sensitivity,  the  mimimum \lya\ line flux is $7 \times 10^{-17} \ {\rm erg/s/cm}^{2}$. We name these the L7 quarter-mocks in what follows.
\\
(b) L3.5 quarter-mocks. Unfortunately, even with the above parameter choices, which are in themselves relatively optimistic regarding the capabilities of RST, we shall see below that the possibility of detecting BAO is marginal. We therefore make another set of quarter-mocks with a 50\% less restrictive line detection sensitivity limit,  $3.5 \times 10^{-17} \ {\rm erg/s/cm}^{2}$. This sensitivity limit could in principle be reached by increasing the integration time and lowering the sky area covered by the HLS. We name these the L3.5 quarter-mocks.
\\
(c) L1 quarter-mocks. The final set of mocks we  use for analysis have a line detection sensitivity of $1 \times 10^{-17} \ {\rm erg/s/cm}^{2}$  7 times better than the planned RST design. We create these (L1) quarter-mocks in order to explore BAO measurement when there is extremely good sampling.

In Fig.~\ref{fig:sfr_dis_mass20to50} we  show galaxies with SFR more than $44.20 \ {\rm yr}^{-1}M_{\sun} $ account for 15.9\% of the total galaxy counts. We assume that if a galaxy mass falls between $2\times 10^{11} \ h^{-1} M_{\sun}$ and $5 \times 10^{11} \ h^{-1} M_{\sun}$, the probability of it being able to be detected is 15.9\%. We calculate this probability for different mass ranges by using data from \textsc{BlueTides}, and match galaxies from simulated cuboid to decide which galaxies will be detected in the mock survey. Fig.~\ref{fig:cross_section} shows the difference after we apply this selection. Only 2\% galaxies is kept in the right figure compared with left one. And, Fig.~\ref{fig:p_mass} shows the relationship between probability and mass. The maximum mass in redshift 9 and 10 is much smaller than other redshift, but considering the maximum mass of the Outer Rim is $1.30 \times 10^{12} \ h^{-1} M_{\sun}$ in redshift 9.2 and $6.7 \times 10^{11} \ h^{-1} M_{\sun}$ in redshift 10. Fig.~\ref{fig:p_mass} is enough to cover the mass range in the Outer Rim. And Fig.~\ref{fig:number_z_L} shows the distribution of galaxies along the z-axis under different line sensitivity. There are several sudden changes in the distribution, the reason is that we use snapshots from discrete redshift to construct mock data so there is discontinuity between two redshift slices.

\subsection{Measuring the correlation function} \label{sec:2pcf}
We use the Landy Szalay estimator \citep[][]{Landy1993} to compute anisotropic two point correlation functions $\hat{\xi}(s)$ of galaxies in our datasets. We prefer the Landy Szalay estimator, since its performance has been proved to be better than other comparable two point correlation functions \citep[][]{DavisPeebles1983, Hamilton1993} at large scales \citep[][]{Pons1999, Kerscher2000}. Equation~\ref{eqn:LSEqn_Anisotropic} gives the formula for the calculation of the Landy Szalay estimator.
    
\begin{equation}
\label{eqn:LSEqn_Anisotropic}
\hat{\xi}_{\rm LS}(s, \mu) = \frac{DD(s, \mu) - 2DR(s, \mu) + RR(s, \mu)}{RR(s, \mu)}.
\end{equation}

In equation~\ref{eqn:LSEqn_Anisotropic}, $\mu=cos \ \theta$ (where $\theta$ is the angle between the line of sight and the pair separation vector for any given pair), $DD(s, \mu)$ depicts the pair count of galaxies with pair separation distance $s$ and orientation $\mu$. The symbol $DR(s, \mu)$ represents the cross-pair counts between galaxies and randoms which have pair separation $s$ and angle $\theta$, and $RR(s, \mu)$ is the number of pairs for a random distribution. We have computed the correlation functions as a function of $s$ and $\mu$ as an additional check on our calculations, to see that the clustering is behaving as expected. We do not present or use the two dimensional results further because the use of the full two-dimensional two point correlation functions in analyses is computationally expensive. One can use isotropised correlation functions ($\tilde{\xi}_{f}(s)$) to circumvent this. To obtain isotropised correlation functions, one uses specific kernels to marginalize out the $\mu$ coordinate in the two point correlation function. In \citet{Hamilton1993} it was shown that one can use Legendre polynomials as a kernel ($f(\mu) = P_{\ell}(\mu) $ to find the isotropic correlation functions $\tilde{\xi}_{\ell}(s)$. Different values of $\ell$ will lead to isotropic correlation functions of different orders. In our research, we use isotropic correlation functions with Legendre polynomials and $\ell=0$ to obtain isotropized monopoles: ($\tilde{\xi}_{0}(s)$). For completeness, Equation \ref{eqn:Isotropized2PCorr} shows the process by which anisotropic correlationfunction ${\xi}_{\ell}(s)$ is obtained from the marginalization of anisotropic two correlation function ($\hat{\xi}(s, \mu)$).

\begin{align}
\label{eqn:Isotropized2PCorr}
\tilde{\xi}_{\ell}(s) &= \frac{2 \ell + 1}{2} \int^{1}_{-1}\hat{\xi}_{\rm LS}(s,\mu) P_{\ell}(\mu)d\mu.
\end{align} 

In equation~\ref{eqn:LSEqn_Isotropic}, we show how we can compute $\tilde{\xi}_{0}(s)$ from equation~\ref{eqn:Isotropized2PCorr}. 

\begin{align}
\label{eqn:LSEqn_Isotropic}
\tilde{\xi}_{0}(s) &= \left(  \frac{1}{2} \right) \int^{1}_{-1}\hat{\xi}_{\rm LS}(s,\mu) d\mu \nonumber \\
&\approx  \left( \frac{1}{2} \right) \displaystyle\sum_{j} \Delta \mu_j \hat{\xi}_{\rm LS}(s,\mu_j).
\end{align}

In the work presented in this paper, we have used 100 bins in $\mu$ in the computation of all two dimensional two-point correlation functions. Also, all the multipoles ($\xi_{0,2}(s)$) that we compute for \textsc{BlueTides} and Outer Rim galaxies have evenly spaced bins of width $3 \ h^{-1}$Mpc in $s$.
Apart from their use as an internal check here, we reserve the presentation of results for multipoles higher than the monopole for future work.

\subsection{Fitting the correlation function} \label{sec:BAOfit}

\begin{figure}
    \includegraphics[width=\columnwidth]{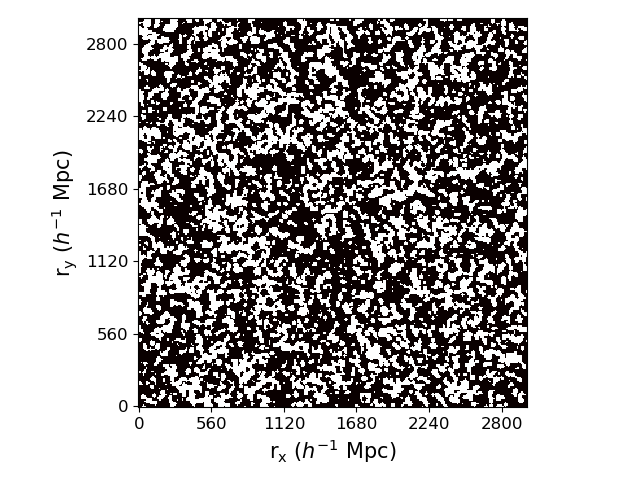}
    \caption{This plot shows a slice at ${\rm r}_{\rm z} = 1500  h^{-1}$ Mpc obtained from the three dimensional grid of cells with ionized and non-ionized regions determined from the reionization redshift field, $z_{\rm RE} (\mathbf{x})$. The slice thicknesses in this three dimensional grid are 10 $h^{-1}$ Mpc. The white patches correspond to regions which are ionized at $z=9.4$, while the black patches denote non-ionized regions.}
    \label{fig:zRE_smooth_selection}
\end{figure}

\begin{figure*}
    \includegraphics[width=0.49\textwidth]{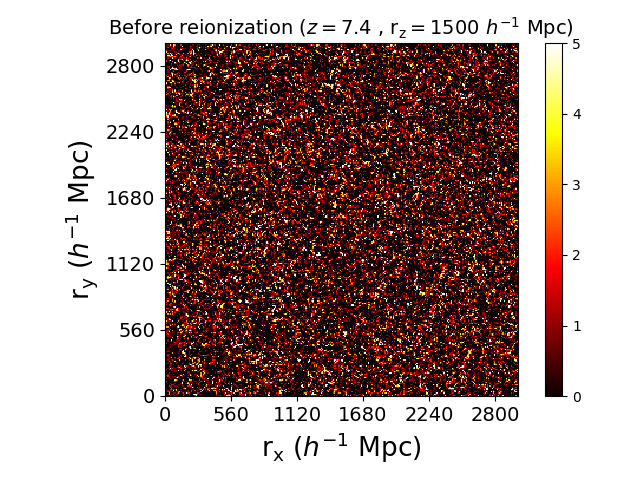}
    \includegraphics[width=0.49\textwidth]{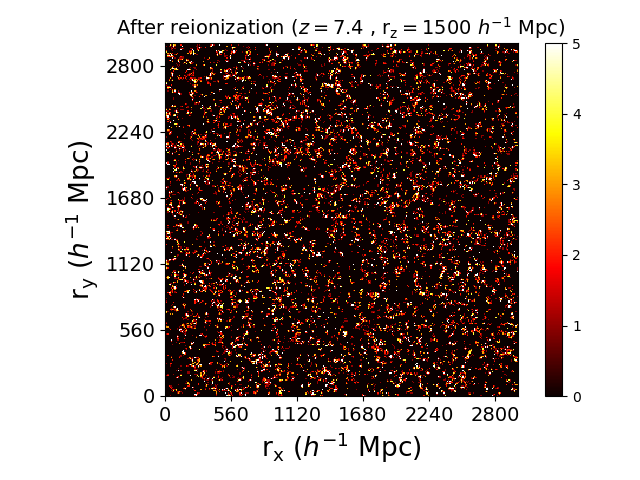}
    \caption{The plot on the left hand side shows a slice (${\rm r}_{\rm z} = 1500  h^{-1}$ Mpc) of Outer Rim data at $z=7.4$ before patchy reionization. In the right hand side, we show a slice (${\rm r}_{\rm z} = 1500  h^{-1}$ Mpc) of  Outer Rim data at $z=7.4$ after reionization. The plot on the right hand side is obtained after Outer Rim data at $z=7.4$ is analyzed using the grid of ionized and non-ionized regions described in section~\ref{sec:reion_model} and illustrated in Fig.~\ref{fig:zRE_smooth_selection}.}   
    \label{fig:before_after_reionization}     
\end{figure*}

For fitting of the correlation functions, we use techniques which are outlined in \citet{Anderson2012, Xu2012, Xu2013, Vargas2014, Ansarinejad2018, Sarpa2019}. We briefly describe the methods that we have used in this section.

We use the following fiducial form to fit space correlation monopoles $\xi_0(s)$ over the range  $33 \ h^{-1} {\rm Mpc} < s < 138 \ h^{-1} {\rm Mpc}$ with bin sizes of $s=3 \ h^{-1}$Mpc.

\begin{align}
\label{eqn:BAO_xi}
\xi^{\rm fit} (s) = B^2 \xi_{ {\rm m} }(\alpha s) + A(s).
\end{align}

Here, the symbol `$s$' represents the pair separation between galaxies in redshift space. The quantity $B$ is a multiplicative constant bias which adjusts the amplitude of the correlation functions ($\xi$), and lets the model account for any unknown large-scale bias.  The term

\begin{align}
\label{eqn:BAO_A}
A(s) = \frac{ a_{1} }{s^2} + \frac{ a_{2} }{s} + a_{3}.
\end{align}
is included so that the fitting model can marginalize over broad-band effects due to scale dependent bias, redshift space distortions and errors during our assumption of the fiducial cosmology. The parameters $ a_{1}, \ a_{2}, \ a_{3} $ in equation~\ref{eqn:BAO_A} are nuisance parameters. The choice of number of parameters in $A(s)$ in our work is inspired by \citet{Anderson2012, Xu2012, Xu2013, Vargas2014, Ansarinejad2018, Sarpa2019}. The inclusion of $A(s)$ can help to mitigate the effects of assumption of a wrong cosmological model. At the same time, $A(s)$ can be also be thought of as an effectual description of mode coupling not affecting the BAO scale but biasing its measurement if not considered properly \citep[][]{Crocce2008}. 

The scale dilation parameter $\alpha$ measures the position of the BAO peak of the data relative to the model. One can also think of the parameter $\alpha$ as the extent to which the acoustic peak in the data is shifted relative to the model. This isotropic shift in the positions of the BAO peak in the data and the model occurs due to non-linear structure growth. A numerical definition of $\alpha$ can be obtained in equation~\ref{eqn:alpha_eqn}.

\begin{align}
\label{eqn:alpha_eqn}
\alpha = \frac{ D_{\rm V} (z) }{ r_d } \frac{ r_{d , {\rm fid}  } }{ D_{ {\rm V} , {\rm fid} } (z) }
\end{align}

In equation~\ref{eqn:alpha_eqn} $r_d$ represents the sound horizon at the drag epoch. The superscript `fid' denotes the fiducial cosmology (given in Table~\ref{tab:cos_bluetides}.) A value of $\alpha<1$ would represent a shift towards larger scales, whereas a value of $\alpha>1$ denotes a shift towards smaller distance scales. The volume averaged distance $D_{\rm V}$ is related to the Hubble constant $H$, the angular diameter distance $D_{\rm A}$, the speed of light $c$ and the redshift $z$ by the following equation:

\begin{equation}
\label{eqn:D_V}
D_{\rm V} = \left[ (1+z)^2 c z \frac{D_{\rm A}^2}{H} \right]^{1/3}
\end{equation}

The template power spectrum which we use is defined in the below mentioned relation. 

\begin{align}
\label{eqn:template_power_spectrum}
P_{\rm m}(k) = \left[ P_{\rm lin}(k)  - P_{\rm smooth} (k)   \right] e^{ -k^2 \Sigma_{ {\rm NL}  }^2 / 2  } + P_{\rm smooth} (k)
\end{align}

In equation~\ref{eqn:template_power_spectrum}, $ P_{\rm lin}(k)$ represents the actual linear power spectrum at $z=0$. The term $P_{\rm smooth} (k)$ denotes the de-wiggled limit where the BAO feature is removed. More details of $P_{\rm smooth} (k)$ can be found in \citet{Eisenstein1998}. The term $\Sigma_{ {\rm NL} }$ helps to damp the BAO features in the linear power spectrum $ P_{\rm lin}(k)$. This parameter also helps to account for the broadening and shift of the BAO feature due to non linear structure evolution. We use $\Sigma_{ {\rm NL} } = 0 \ h^{-1}$Mpc in our work, i.e. $P_{ {\rm m} }(k) = P_{\rm lin}(k)$. Choice of this value of $\Sigma_{ {\rm NL} }$ is justified given the high redshifts that we consider ($z_{\rm OuterRim}=9.4$) and the correspondingly low value of the growth factor, $g=0.123$. More details of this are given in Section~\ref{sec:reion_model}. 

We use \textsc{CAMB} \citep[][]{Lewis2002} to compute the power spectrum $P_{ {\rm m} }(k)$ using the same cosmological model that is adopted in the Outer Rim simulation.

The model correlation function $\xi_{\rm m}$ is the Fourier transform of the the template power spectrum given in equation~\ref{eqn:template_power_spectrum}. 

\cite{Xu2013} showed that variations in the bias like term $B$ do not affect the position of the BAO peak in monopoles. At the same time, the parameter $\alpha$, which gives us a quantification of the acoustic scale, affects the position of the BAO peak. As such, $\alpha$ is the parameter that we are most interested in obtaining from our fits. The models that we use are non-linear in $\alpha$, so we can nest a linear least-squares fitter inside a non-linear fitting routine. The best fit value of $\alpha$ is obtained by minimizing the $\chi^2$ goodness-of-fit indicator (non-linear fitter). To obtain an optimum value of $\alpha$, we compute $\chi^2$ and shift models in the range $0.8 < \alpha < 1.2$ with intervals of $\alpha=0.01$. The choice of range within which values of $\alpha$ are examined, \textit{viz.} $\left[ 0.8, 1.2 \right]$, is inspired by \citet{Vargas2014} and \citet{Ansarinejad2018}. 

The best fit value of $\alpha$ is obtained by minimizing the $\chi^2$ goodness-of-fit indicator given in equation~\ref{eqn:BAO_chi2}.

\begin{align}
\label{eqn:BAO_chi2}
\chi^2(\alpha) = \lbrack \xi^{\rm mock} - \xi^{\rm fit} (\alpha) \rbrack ^{\rm T}  \ \mathbf{\Sigma}^{-1} \  \lbrack \xi^{\rm mock} - \xi^{\rm fit} (\alpha) \rbrack
\end{align}

In equation~\ref{eqn:BAO_chi2}, $ \xi^{\rm fit}(\alpha)$ represents a model vector at a given value of $\alpha$, $\xi^{\rm mock}$ denotes the correlation function measured from mock cuboids corresponding to a given line detection sensitivity, and $\mathbf{\Sigma}$ represents a diagonal covariance matrix which is obtained from all mock cuboids in the considered line detection sensitivity. In an ideal case, one would have prefered to use a variance-covariance matrix from a large number (preferably $\sim$100 or higher) of mock catalogues. But, in our work, we are dealing with high redshifts ($z>7.6$) and we only have four mock cuboids for each of the three line detection sensitivities. We use these cuboids to construct separate covariance matrices for each line detection sensitivity. Hereafter, we shall use the symbol $\alpha^{\ast}$ to denote the best fit values of the scale dilation parameters $\alpha$. To obtain robust results for best fit values of scale dilation parameters and best fit model vectors ($\xi^{\rm fit} (\alpha^{\ast}) $), we construct diagonal covariance matrices from mocks in two steps. In the first step, we compute a diagonal covariance matrix (say $\mathbf{\Sigma}_0$) from cuboids which correspond to the same line detection sensitivity. (Separate $\mathbf{\Sigma}_0$ are constructed for catalogues which correspond to without patchy reionization and with patchy reionization.) In the second step, we construct a new diagonal covariance matrix (say $\mathbf{\Sigma}$) by using a moving average (with a window size of three) over the diagonal elements of $\mathbf{\Sigma}_0$. That is to say,

\begin{align}
\label{eqn:covmat_slidingwindow1}
\mathbf{\Sigma}^{ \lbrace i , i \rbrace } = \frac{ \left[ \mathbf{\Sigma}_0^{ \lbrace (i-1) , (i-1) \rbrace  } + \mathbf{\Sigma}_0^{ \lbrace i , i \rbrace  } + \mathbf{\Sigma}_0^{ \lbrace (i+1) , (i+1) \rbrace  } \right] }{3}
\end{align}

To compute $\mathbf{\Sigma}^{ \lbrace 0 , 0 \rbrace }$, we use equation~\ref{eqn:covmat_slidingwindow2}

\begin{align}
\label{eqn:covmat_slidingwindow2}
\mathbf{\Sigma}^{ \lbrace 0 , 0 \rbrace } = \frac{ \left[ \mathbf{\Sigma}_0^{ \lbrace 0 , 0 \rbrace  } + \mathbf{\Sigma}_0^{ \lbrace 1 , 1 \rbrace  } \right] }{2}
\end{align}

Similarly, to find the last diagonal element of $\mathbf{\Sigma}$, i.e. $\mathbf{\Sigma}^{ \lbrace N , N \rbrace }$ we use equation~\ref{eqn:covmat_slidingwindow3}. Here $N$ denotes the index of the largest distance scale ($138 \ h^{-1} {\rm Mpc}$) in our fitting range.

\begin{align}
\label{eqn:covmat_slidingwindow3}
\mathbf{\Sigma}^{ \lbrace N , N \rbrace } = \frac{ \left[ \mathbf{\Sigma}_0^{ \lbrace (N-1) , (N-1) \rbrace  } + \mathbf{\Sigma}_0^{ \lbrace N  , N \rbrace  } \right] }{2}
\end{align}

The approach highlighted in equations~\ref{eqn:covmat_slidingwindow1} ,~\ref{eqn:covmat_slidingwindow2} and~\ref{eqn:covmat_slidingwindow3} help in getting better agreement in positions of BAO peaks in mock and best fit model correlation functions.

Detailed description of the BAO fitting methodologies can also be found in \citet{ Anderson2012, Xu2012, Xu2013, Vargas2014}.

\subsection{Semi analytic model for patchy reionization} \label{sec:reion_model}

We follow the semi-analytic model outlined in \cite{Battaglia2013} to construct reionization fields from our density fields. The density fluctuation field is supposed to be at the midpoint of reionization, which is $z=9.5$ for \textsc{BlueTides}. So, we use data from Outer Rim simulation which is closest to $z=9.5$ (i.e. $z_{\rm OuterRim}=9.4$). The fluctuation field for halos from the Outer Rim catalogue at $z_{\rm OuterRim}=9.4$ is defined as:

\begin{align}
\label{eqn:halo_fluc_field}
\delta_{\rm halo} (\mathbf{x}) \equiv \frac{  \delta_{\rm halo}(\mathbf{x}) - \bar{\delta}_{\rm halo}   }{ \bar{\delta}_{\rm halo}  }
\end{align}

Here, $\bar{\delta}_{\rm halo}$ is the mean density of halos. The required linear bias, $b$ is obtained by using equation~\ref{eqn:linear_bias}.

\begin{align}
\label{eqn:linear_bias}
b = \frac{\sigma_8}{g} \cdot \sqrt{ \frac{ \xi_0^{z=9.4 \ ; \ {\rm OuterRim}} }{  \xi_0^{z=0.0 \ ; \ {\rm CDM}} } }
\end{align}

In equation~\ref{eqn:linear_bias}, $g$ is the growth factor between $z=9.4$ and $z=0.0$, i.e. $g=0.123$. The symbol $\xi_0^{z=9.4 \ ; \ {\rm OuterRim}}$ denotes the correlation function monopole computed from Outer Rim halos at $z=9.4$. The symbol $\xi_0^{z=0.0 \ ; \ {\rm CDM}}$ represents the monopole from cold dark matter (CDM) at $z=0.0$. We use a value of $\sigma_8 = 0.8$.
After the aforementioned steps, one can obtain the matter overdensity ($\delta_{\rm m}$) from equation~\ref{eqn:matter_overdensity}

\begin{align}
\label{eqn:matter_overdensity}
\delta_{\rm m} (\mathbf{x}) = \delta_{\rm halo} (\mathbf{x}) / b
\end{align}

Before we outline the steps for the computation of the reionization redshift field, $z_{\rm RE}(\mathbf{x})$, it is important to define the fluctuation field $\delta_z(\mathbf{x})$.

\begin{align}
\label{eqn:overdensityz}
\delta_{\rm z}(\mathbf{x}) \equiv \frac{  \left[ 1 + z_{\rm RE}(\mathbf{x}) \right] - \left[ 1 + \bar{z} \right] }{1 + \bar{z}}
\end{align}

In equation~\ref{eqn:overdensityz}, $\bar{z}$ is the mean value of the $z_{\rm RE}$ field. At the same time, $\bar{z}$ sets the midpoint of reionization which is $z=9.4$ for Outer Rim data. 

From here, we can get the reionization redshift field, $z_{\rm RE} (\mathbf{x})$, by following the below mentioned steps.

\begin{enumerate}
\item[1.] We Fourier transform the density fluctuation field $\delta_{\rm m}$ from Outer Rim simulation at the midpoint of reionization (i.e. $z=\bar{z}=9.4$) and obtain $\tilde{ \delta }_{\rm m} (k)$.
\item[2.] We multiply the field $\tilde{ \delta }_{\rm m} (k)$ with the bias factor given in equation~\ref{eqn:bias_factor} to obtain $\tilde{ \delta }_{\rm z} (k)$.

\begin{align}
\label{eqn:bias_factor}
b_{\rm mz}(k) = \frac{b_0}{ \left( 1 + k/k_0 \right)^{\alpha} }
\end{align}

The above mentioned bias factor uses three parameters, \textit{viz.} $b_0$, which is the bias amplitude on the largest scales, $k_0$, which is the scale threshold and $\alpha$, which is the asymptotic exponent. For the three parameters $b_0$, $k_0$, $\alpha$, we use the best-fit values which were used in \citet{Battaglia2013}. More specifically, we use values of $b_0 = 0.593$ , $k_0 = 0.185 \ {\rm Mpc}^{-1} h$ and $\alpha = 0.564$ in our work. Since we don't focus much on small distance scales in our work, we ignore the top hat filters that are used in \citet{Battaglia2013}.

\item[3.] We inverse Fourier transform $\tilde{ \delta }_{\rm z} (k)$ to real space to get the $\delta_{\rm z}$ field.

\item[4.] Now, we can convert the $\delta_{\rm z}$ field to $z_{\rm RE}$ using equation~\ref{eqn:overdensityz}.

\end{enumerate}

Once we have the $z_{\rm RE} (\mathbf{x})$ field, we can use that to modulate the galaxy distribution. Basically, we can use the same grid for all snapshots, and if a galaxy is in a cell that has reionized by that snapshot redshift, we treat it as being visible. If a galaxy is in a cell which hasn't reionized (these will be the lower density cells, based on how the patchy reionization model works), then we say that it won't be visible, and we remove it from the sample. As with the previous mock catalogues, we include the effect of redshift-space distortions in the galaxy positions: we use real space position to locate the galaxy in the reionization field, and then position the visible galaxies
in redshift space.

We follow the following prescription to distinguish between lower and higher density cells. We use a Gaussian filter with $\sigma=10 \ h^{-1}$Mpc to smooth the $z_{\rm RE} (\mathbf{x})$ field. In the smoothed $z_{\rm RE} (\mathbf{x})$ field, any cell whose value is less than the mean value of the smoothed $z_{\rm RE} (\mathbf{x})$ field is considered as a low density cell. Similarly, any cell in the the smoothed $z_{\rm RE} (\mathbf{x})$ field whose value is bigger than the mean is considered as a higher density cell. Fig.~\ref{fig:zRE_smooth_selection} shows a slice (${\rm r}_{\rm z} = 1500  h^{-1}$ Mpc) of the three dimensional grid of cells with ionized and non-ionized regions which is obtained from the aforesaid method. We compare all snapshots (redshifts) with this grid of high and low density cells and construct a new catalogue of galaxies which correspond to scenarios after patchy reionization. In Fig.~\ref{fig:before_after_reionization}, we see slices (at ${\rm r}_{\rm z} = 1500  h^{-1}$ Mpc) from Outer Rim data at $z=7.4$ with and without patchy reionization.

We measure correlation function monopoles in these galaxy catalogues which correspond the cases with and without patchy reionization. In the non-patchy case, we assume that the intervening intergalactic medium does not affect the visibility of the \lya\ line from galaxies (i.e.  the universe is optically thin at the time of \lya\ emission).Results of monopoles and the ensuing BAO fitting are presented in Section~\ref{sec:bao_results}.

\section{Results} \label{sec:bao_results}

We show the fitting results of correlation functions from L7, L3.5 and L1 quarter mocks for scenarios with and without patchy reionization.  The L7, L3.5 and L1 represent different observational sensitivities (with L7 being the fiducial RST HLS choice.) For each quarter mock (and for each scenario), we have results for the variation of $\chi^2$ with scale dilation parameter $\alpha$ for the four cuboids. At the same time, we show monopoles computed from the Landy Szalay estimator and best fit monopoles for each cuboid in each quarter mock (for each of the two aforementioned scenarios). The best fit monopoles correspond to the best fit value of $\alpha$ (i.e. $\alpha^{\ast}$). The error on the angular diameter distance to the redshift $z=7.7$ Universe is given by the error on the mean of the
$\alpha^{\ast}$ values from the four quarter-mocks.

\subsection{L7 mocks}

\begin{figure}
    \includegraphics[width=0.48\textwidth]{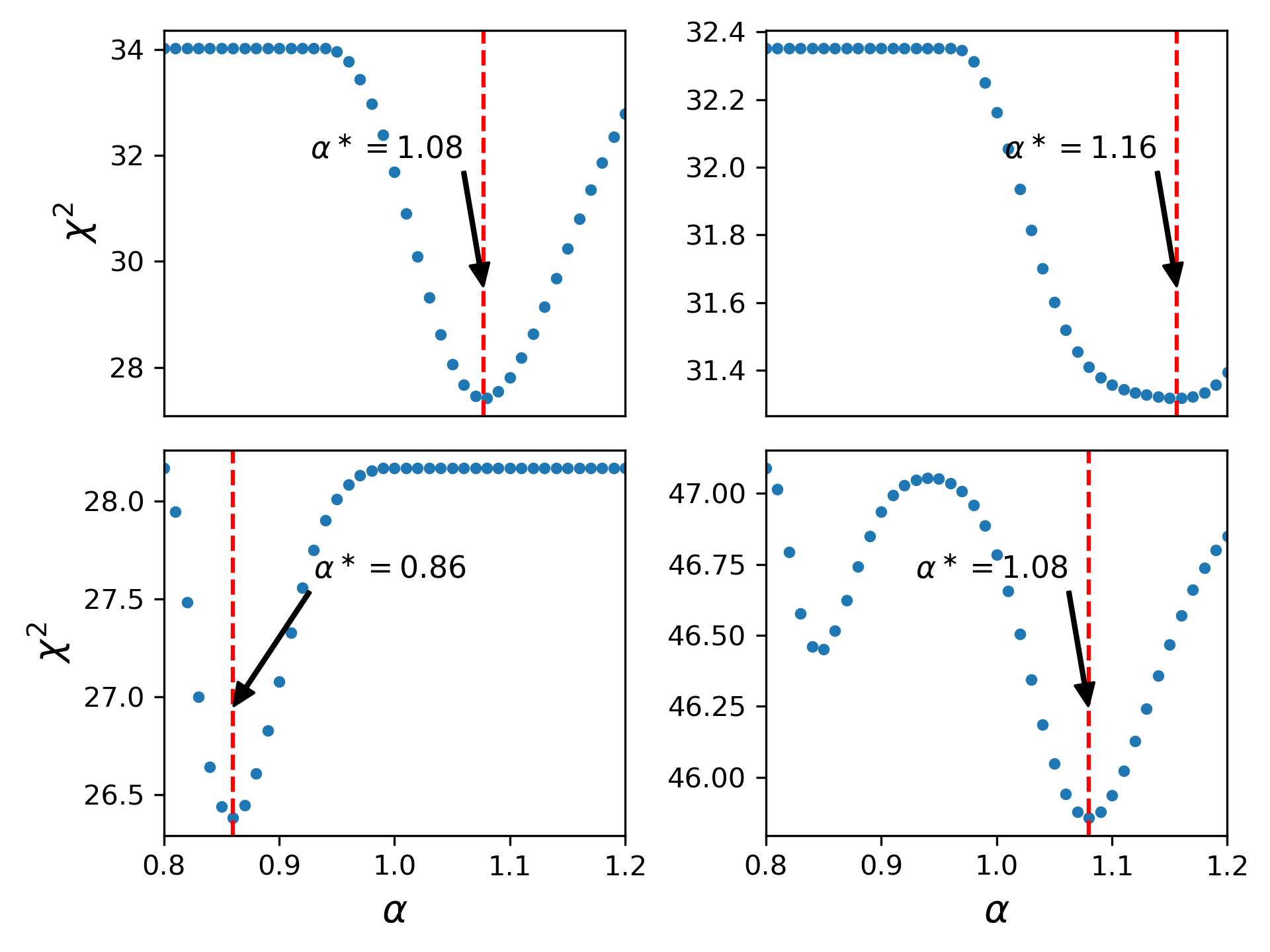}
    \includegraphics[width=0.48\textwidth]{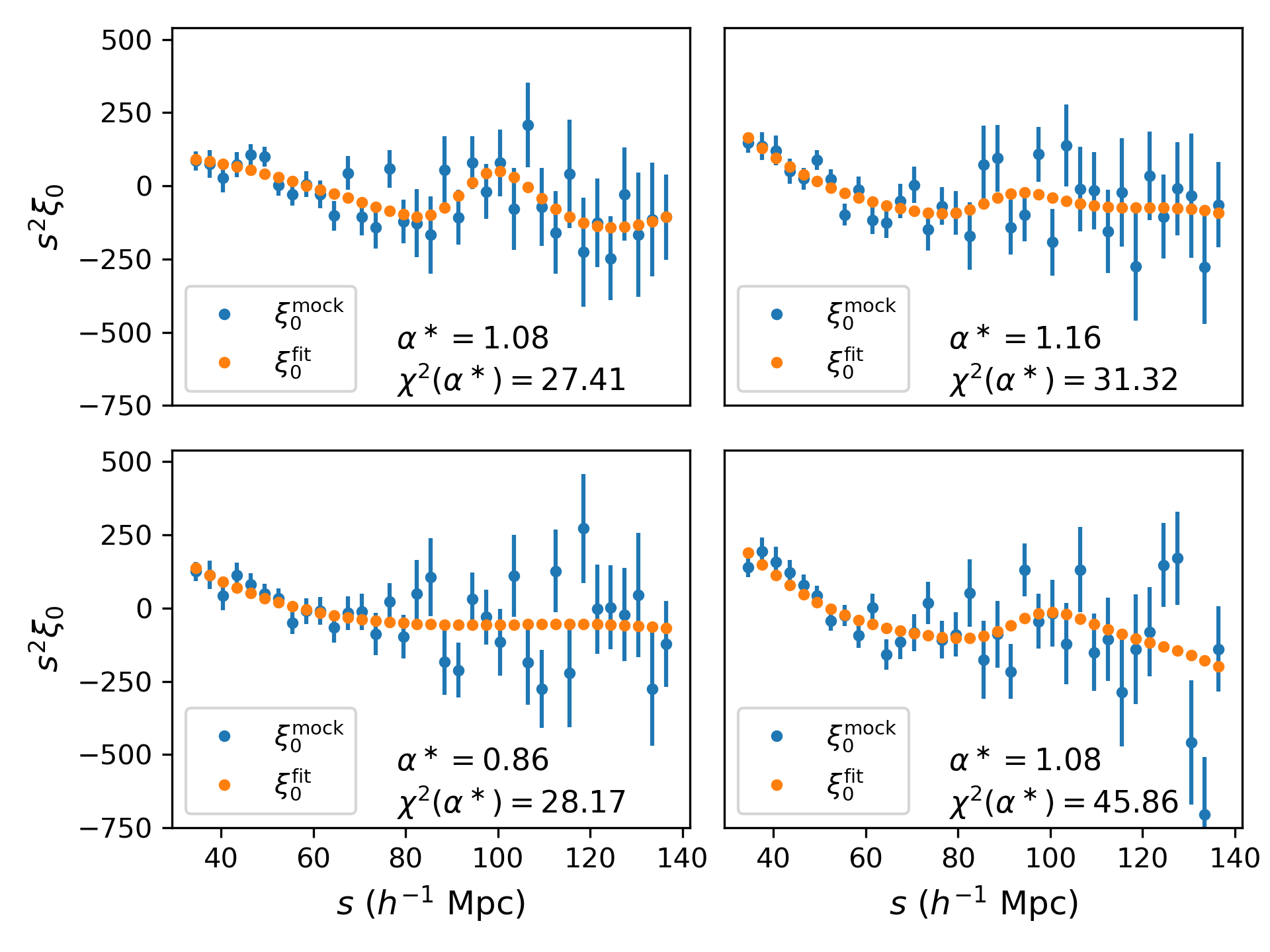}
    \caption{BAO fitting without patchy reionization: In the top two panels we show plots of the variations of $\chi^2$ with $\alpha$ for the four mock cuboids corresponding to $L = 7\times 10^{-17} \ {\rm erg/s/cm}^{2}$ (without patchy reionization). In the plots in the top four panels, the red dashed vertical lines denote the positions of the best fit $\alpha$, \textit{viz.} $\alpha^{\ast}$ for different mocks. In the figures in the bottom four panels, we show the monopoles computed from Landy Szalay estimator ($\xi_0^{\rm mock}$) and the best fit monopoles ($\xi_0^{\rm fit}(\alpha^{\ast})$) obtained after fitting of $\xi_0^{\rm mock}$ for the four mock cuboids corresponding to $L = 7\times 10^{-17} \ {\rm erg/s/cm}^{2}$ (without patchy reionization). In the plots in the bottom panels, the blue dots represent $\xi_0^{\rm mock}$ while the orange dots denote $\xi_0^{\rm fit}(\alpha^{\ast})$.}
    \label{fig:alpha_L0.5_no_reion}
\end{figure}

The top four panels in Fig.~\ref{fig:alpha_L0.5_no_reion} show plots of variation of $\chi^2$ with $\alpha$ for the four mock cuboids in L7 quarter-mock catalogue which correspond to distribution of halos without patchy reionization. The plots in the bottom four panels of Fig~\ref{fig:alpha_L0.5_no_reion} show monopoles computed from Landy Szalay estimators and best fit monopoles ($\xi (\alpha^{\ast})$) for the aforementioned mock cuboids. For the cuboids which correspond to distribution of halos without patchy reionization, the minimum values of $\chi^2$ are 27.41, 31.32, 28.17, 45.86. For these cuboids, the best fit vaues of $\alpha$ are 1.08, 1.16, 0.86, 1.08 respectively. Consequently, for the cuboids from L7 quarter-mocks the mean value of minimum $\chi^2$ is 33.19, and the mean value of $\alpha^{\ast}$ is 1.04. We estimate the fractional error on the distance to $z=7.8$ (the mean galaxy redshift) by computing the sample standard deviation of the  $\alpha^{\ast}$ results for the four quarter-mocks, and then dividing by $\sqrt{4}$ to give the error on the mean. The fractional error computed this way is 0.064, i.e., a distance error accurate to $6.4\%$. We can see from the bottom panel of Fig.~\ref{fig:alpha_L0.5_no_reion}  however that one of the BAO peaks is not visibly detected.

\begin{figure*}
    \includegraphics[width=0.48\textwidth]{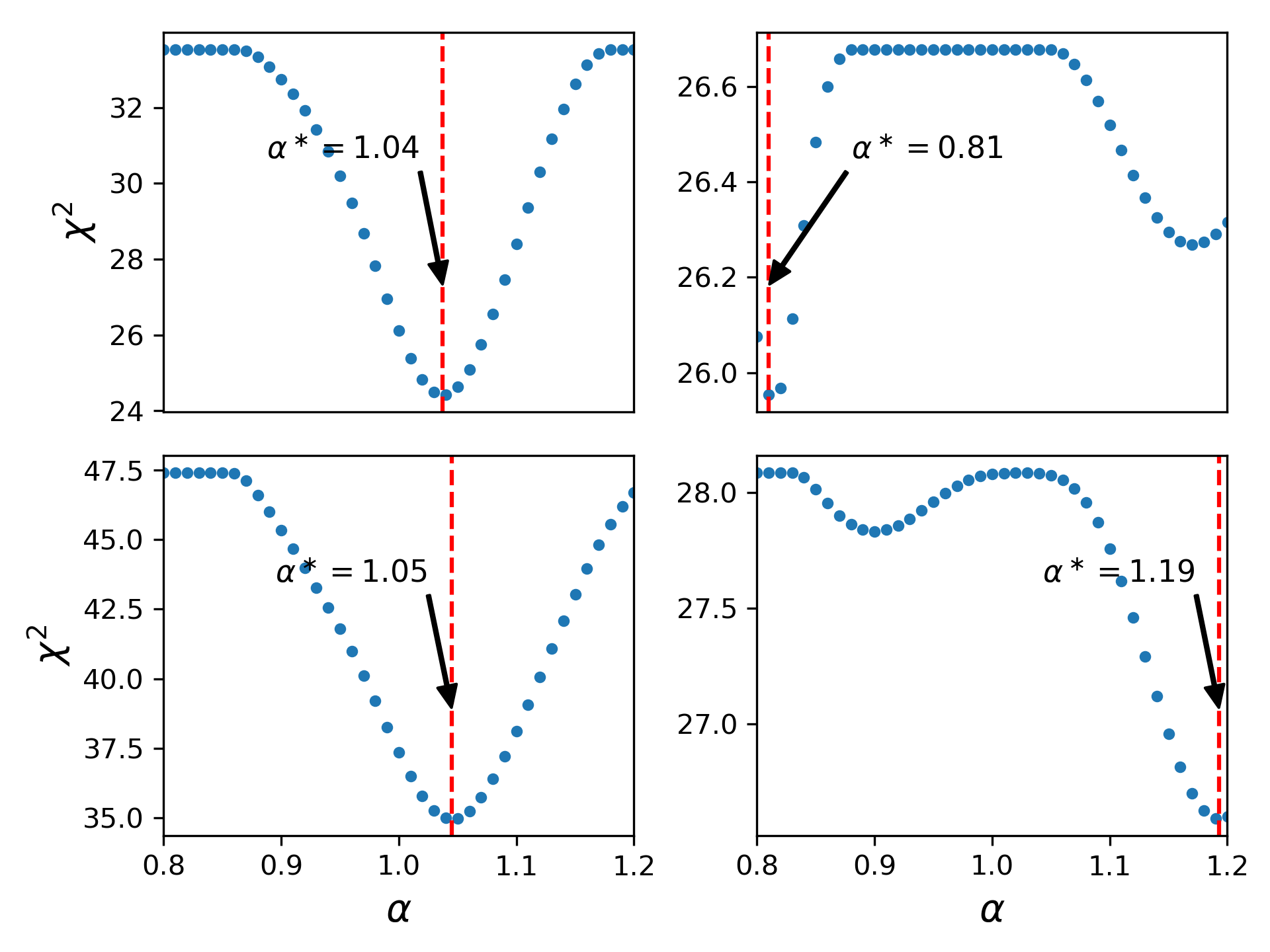}
    \includegraphics[width=0.48\textwidth]{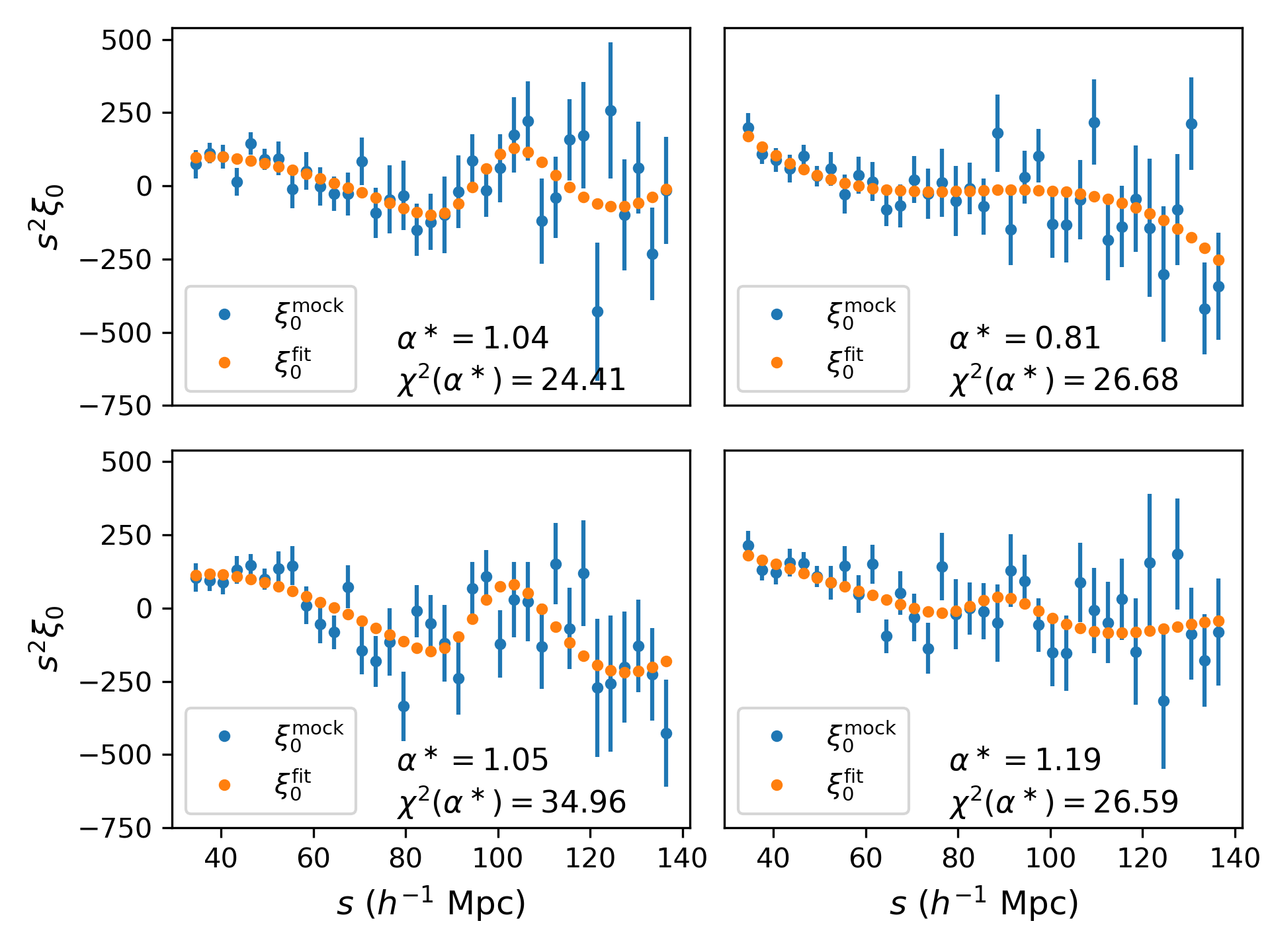}
    \caption{BAO fitting with patchy reionization: In the left four panels of this figure we show the variations of $\chi^2$ with $\alpha$ for the four mock cuboids corresponding to $L = 7\times 10^{-17} \ {\rm erg/s/cm}^{2}$ (after reionization). In the plots in the left four panels, the red dashed vertical lines denote the positions of the best fit $\alpha$, \textit{viz.} $\alpha^{\ast}$ for different mocks. In the right four panels , we show the monopoles computed from Landy Szalay estimator ($\xi_0^{\rm mock}$) and the best fit monopoles ($\xi_0^{\rm fit}(\alpha^{\ast})$) obtained after fitting of $\xi_0^{\rm mock}$ for the four mock cuboids corresponding to $L = 7\times 10^{-17} \ {\rm erg/s/cm}^{2}$ (with patchy reionization). In all the plots in the right panels, the blue dots represent $\xi_0^{\rm mock}$ while the orange dots denote $\xi_0^{\rm fit}(\alpha^{\ast})$. }
    \label{fig:alpha_L0.5_post_reion}
\end{figure*}

One thing that is worth noting as this point is the fact that for a given mock catalogue (L7 or L3.5 or  L1), plots at corresponding positions in the figures which illustrate results with and without patchy reionization are results gotten from the same cuboid. That is, if the leftmost and topmost subplot of Fig.~\ref{fig:alpha_L0.5_no_reion} corresponds to Cuboid 1, then leftmost and topmost subplot of Fig.~\ref{fig:alpha_L0.5_post_reion} also corresponds to Cuboid 1. The same holds of all mock catalogues (L7 or L3.5 or  L1).

The left four panels in Fig.~\ref{fig:alpha_L0.5_post_reion} illustrate the variations of $\chi^2$ with $\alpha$ for the four mock cuboids in L7 quarter-mock catalogue which correspond to distribution of halos after applying patchy reionization. In the right four panels in Fig.~\ref{fig:alpha_L0.5_post_reion}, we show monopoles computed from Landy Szalay estimators and best fit monopoles ($\xi (\alpha^{\ast})$) for these four mock cuboids. The cuboids from L7 quarter-mock catalogue which correspond to distribution of halos with patchy reionization have the following minimum $\chi^2$ values: 24.41, 26.68, 34.96, 26.59. These cuboids have best fit vaues of $\alpha$ as 1.04, 0.81, 1.05, 1.19 respectively. As such, the mean value of minimum $\chi^2$ for these cuboids is 28.16, and the mean value of $\alpha^{\ast}$ is 1.02. Again, as for the previous figure, one of the BAO peaks (in the top right panel) is not detected (although there is a chi-squared minimum and an estimate of  $\alpha^{\ast}$ ). From the scatter of  $\alpha^{\ast}$ values between quarter-mocks, we find the fractional error on the mean distance to be 0.079.

\subsection{L3.5 mocks}

\begin{figure*}
    \includegraphics[width=0.48\textwidth]{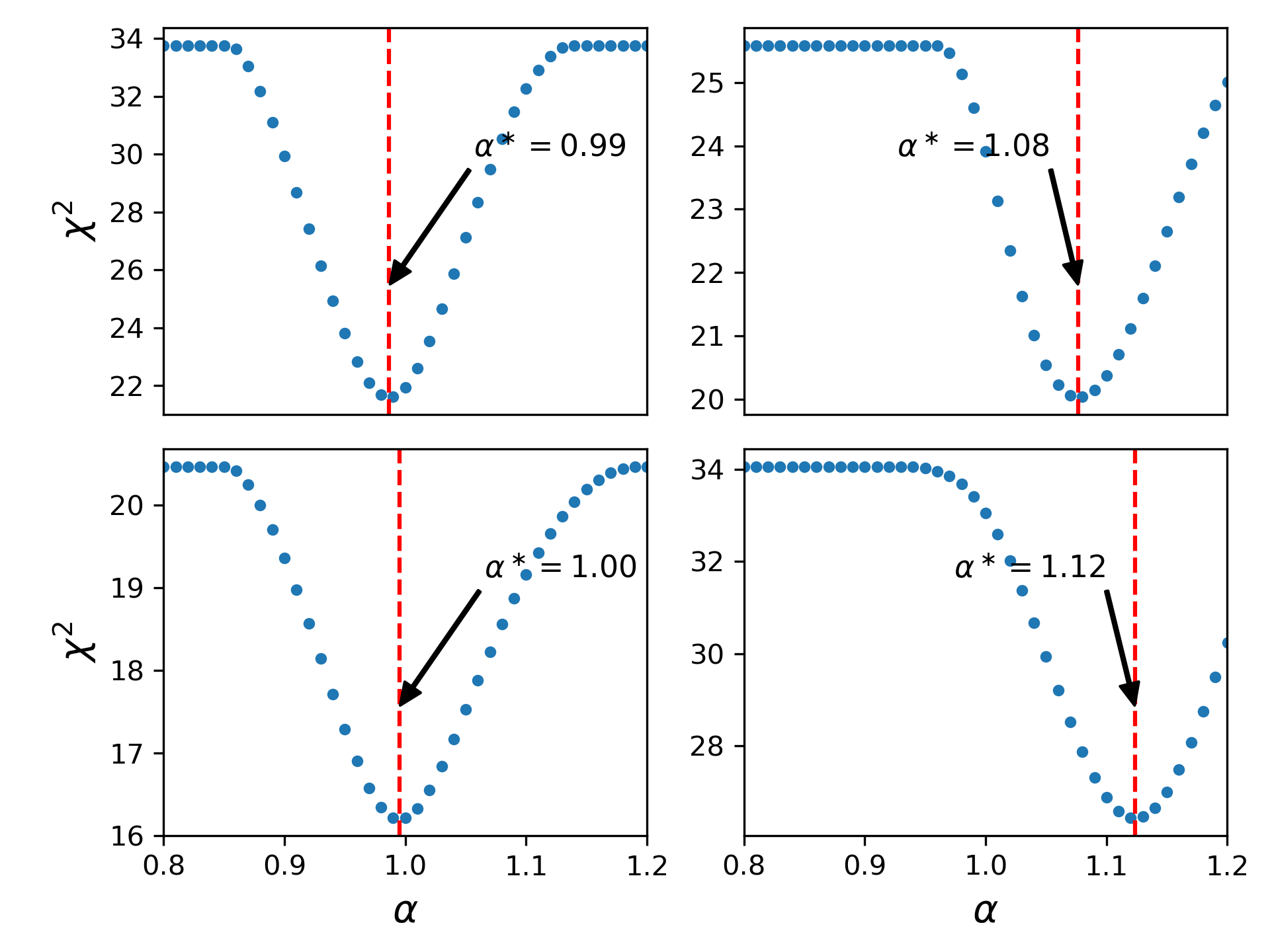}
    \includegraphics[width=0.48\textwidth]{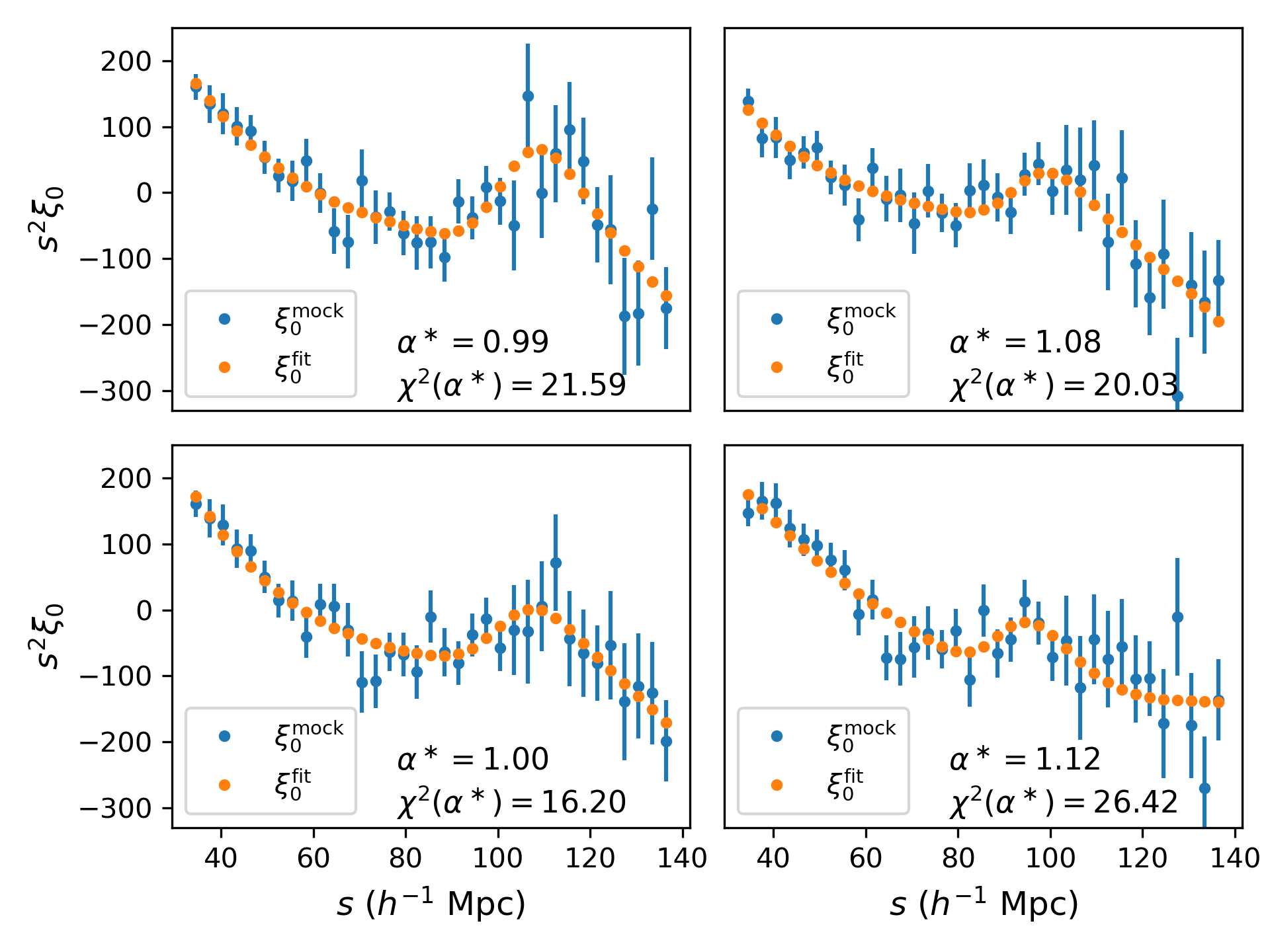}    
    \caption{In the left four panels of this figure we show the variations of $\chi^2$ with $\alpha$ for the four mock cuboids corresponding to $L = 3.5\times 10^{-17} \ {\rm erg/s/cm}^{2}$ (without patchy reionization). In the plots in the left four panels, the red dashed vertical lines denote the positions of the best fit $\alpha$, \textit{viz.} $\alpha^{\ast}$ for different mocks. In the plots in the right four panels, we show the monopoles computed from Landy Szalay estimator ($\xi_0^{\rm mock}$) and the best fit monopoles ($\xi_0^{\rm fit}(\alpha^{\ast})$) obtained after fitting of $\xi_0^{\rm mock}$ for the four mock cuboids corresponding to $L = 3.5\times 10^{-17} \ {\rm erg/s/cm}^{2}$ (without patchy reionization). In each plot in the right panels, the blue dots represent $\xi_0^{\rm mock}$ while the orange dots denote $\xi_0^{\rm fit}(\alpha^{\ast})$. }
    \label{fig:alpha_L0.3_no_reion}
\end{figure*}

The left four panels of Fig.~\ref{fig:alpha_L0.3_no_reion} show plots of variation of $\chi^2$ with $\alpha$ for the four mock cuboids in L3.5 quarter-mock catalogue which correspond to distribution of halos without patchy reionization. In the right four panels of Fig~\ref{fig:alpha_L0.3_no_reion}, we show monopoles computed from Landy Szalay estimators and best fit monopoles ($\xi (\alpha^{\ast})$) for the aforementioned mock cuboids. For the cuboids from L3.5 quarter-mock which correspond to the distribution of halos without patchy reionization, the values of minimum $\chi^2$ are 21.59, 20.03, 16.20, 26.42, and the values of $\alpha^{\ast}$ are 0.99, 1.08, 1.00, 1.12 respectively. Hence, the mean values of minimum $\chi^2$ and $\alpha^{\ast}$ are 21.06 and 1.05 respectively. The error on the BAO distance measurement is 0.032.

\begin{figure*}
    \includegraphics[width=0.48\textwidth]{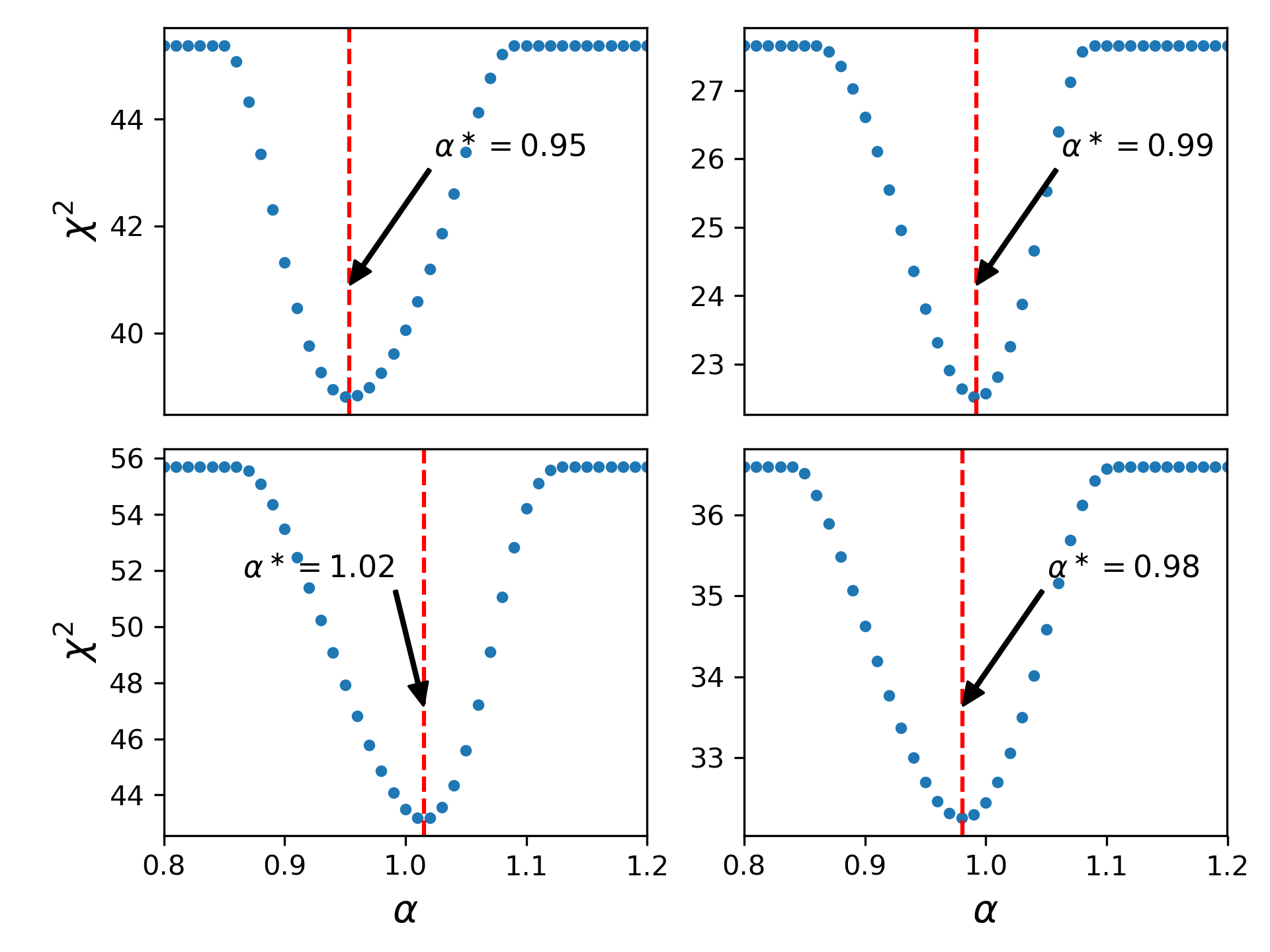}
    \includegraphics[width=0.48\textwidth]{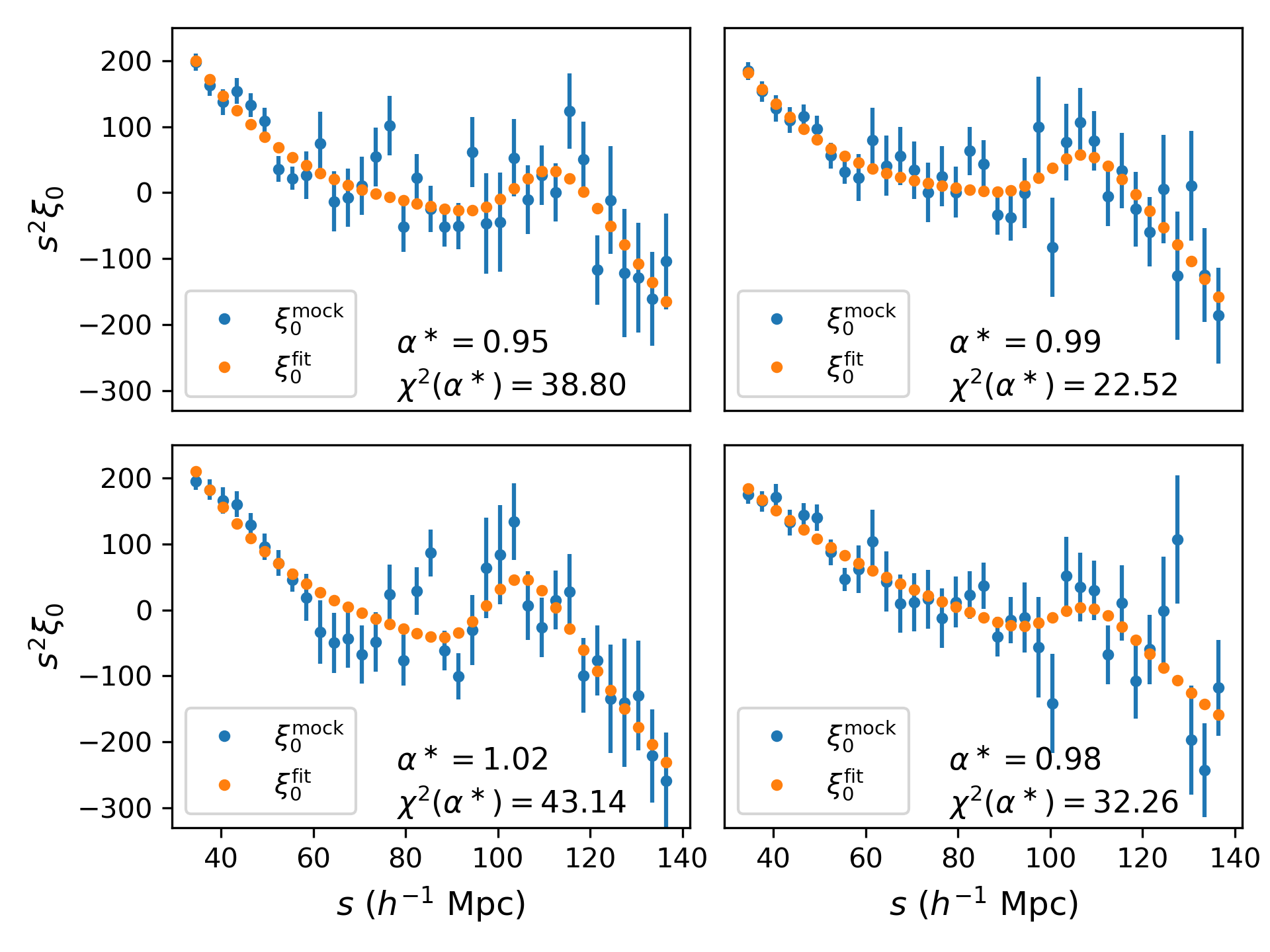}
    \caption{In the left four panels of this figure we show the variations of $\chi^2$ with $\alpha$ for the four mock cuboids corresponding to $L = 3.5\times 10^{-17} \ {\rm erg/s/cm}^{2}$ (with patchy reionization). In the plots in the left four panels, the red dashed vertical lines denote the positions of the best fit $\alpha$, \textit{viz.} $\alpha^{\ast}$ for different mocks. In the right four panels, we show the monopoles computed from Landy Szalay estimator ($\xi_0^{\rm mock}$) and the best fit monopoles ($\xi_0^{\rm fit}(\alpha^{\ast})$) obtained after fitting of $\xi_0^{\rm mock}$ for the four mock cuboids corresponding to $L = 3.5\times 10^{-17} \ {\rm erg/s/cm}^{2}$ (with patchy reionization). In the plots in the right four panels, the blue dots represent $\xi_0^{\rm mock}$ while the orange dots denote $\xi_0^{\rm fit}(\alpha^{\ast})$.}
    \label{fig:alpha_L0.3_post_reion}
\end{figure*}

The plots in the left four panels of Fig.~\ref{fig:alpha_L0.3_post_reion} illustrate the variations of $\chi^2$ with $\alpha$ for the four mock cuboids in L3.5 quarter-mock catalogue which correspond to distribution of halos after applying patchy reionization. In the bottom two rows of Fig.~\ref{fig:alpha_L0.3_post_reion}, we show monopoles computed from Landy Szalay estimators and best fit monopoles ($\xi (\alpha^{\ast})$) for these four mock cuboids. These cuboids have minimum values of $\chi^2$ of 38.80, 22.52, 43.14, 32.26 and best fit values of $\alpha$ as 0.95, 0.99, 1.02, 0.98. As such, the mean values of minimum $\chi^2$ and $\alpha^{\ast}$ are 34.18 and 0.98 respectively. 
The error on the BAO distance measurement is 0.014.

\subsection{L1 mocks}

\begin{figure*}
    \includegraphics[width=0.48\textwidth]{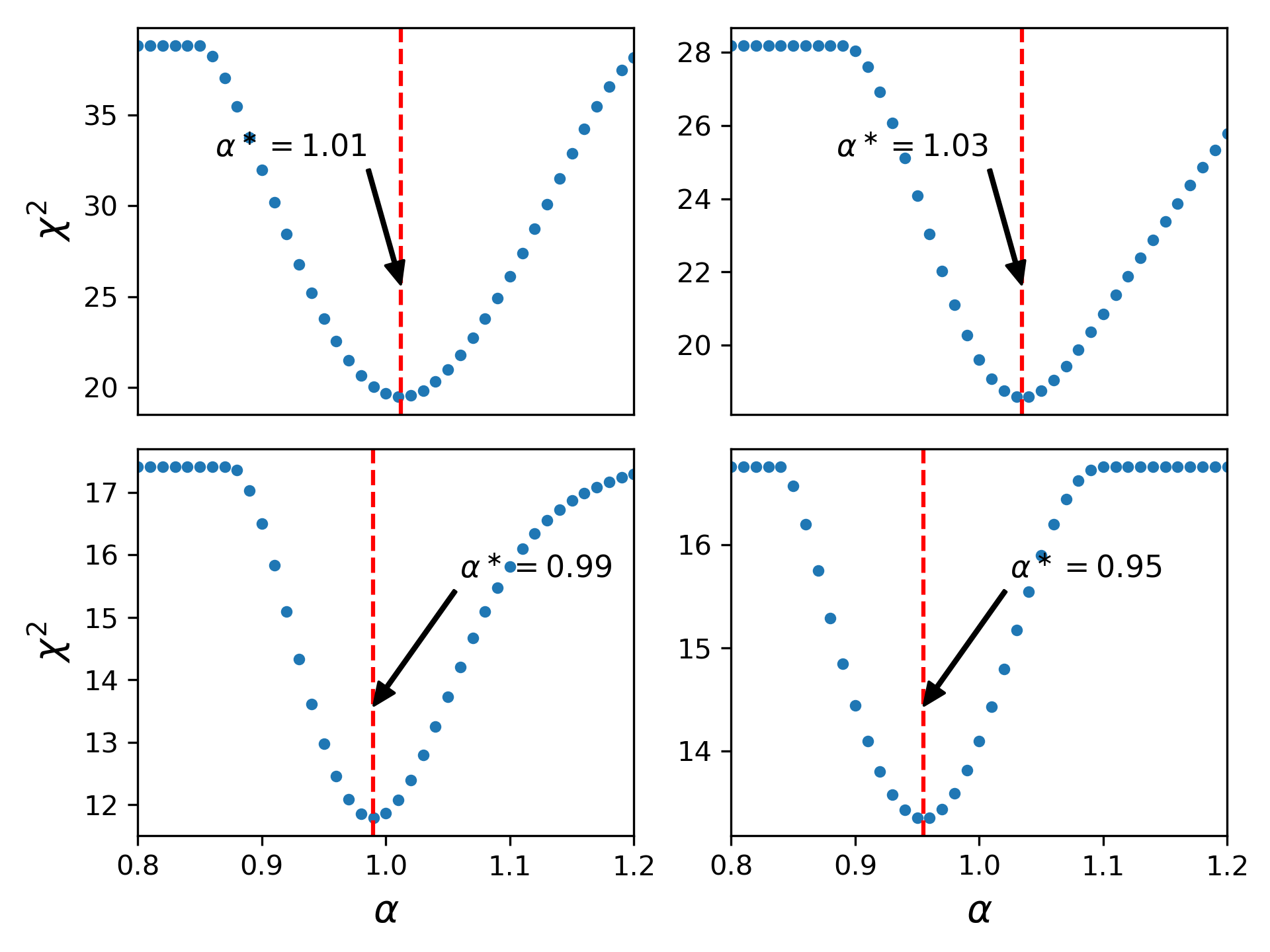}
    \includegraphics[width=0.48\textwidth]{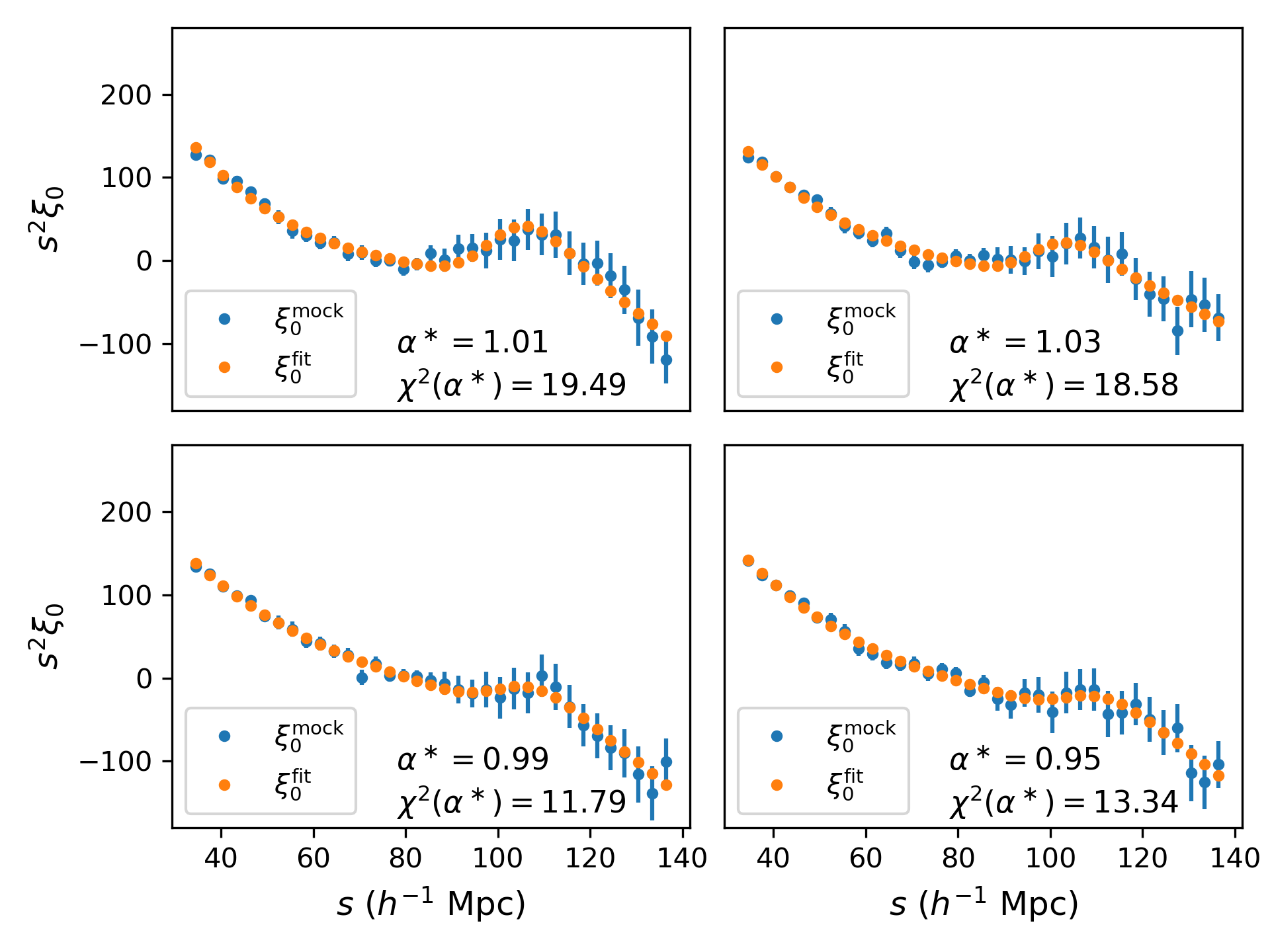}
    \caption{In the left four panels of this figure we show the variations of $\chi^2$ with $\alpha$ for the four mock cuboids corresponding to $L = 1\times 10^{-17} \ {\rm erg/s/cm}^{2}$ (without patchy reionization). In the plots in the left four panels, the red dashed vertical lines denote the positions of the best fit $\alpha$, \textit{viz.} $\alpha^{\ast}$ for different mocks. In the right panels, we show the monopoles computed from Landy Szalay estimator ($\xi_0^{\rm mock}$) and the best fit monopoles ($\xi_0^{\rm fit}(\alpha^{\ast})$) obtained after fitting of $\xi_0^{\rm mock}$ for the four mock cuboids corresponding to $L = 1\times 10^{-17} \ {\rm erg/s/cm}^{2}$ (without patchy reionization). In the plots in the right panels, the blue dots represent $\xi_0^{\rm mock}$ while the orange dots denote $\xi_0^{\rm fit}(\alpha^{\ast})$.}
    \label{fig:alpha_L0.1_no_reion}
\end{figure*}

The left four panels of Fig.~\ref{fig:alpha_L0.1_no_reion} show plots of variations of $\chi^2$ with $\alpha$ for the four mock cuboids in L1 quarter-mock catalogue which correspond to distribution of galaxies visible without patchy reionization. In the right four panels of Fig.~\ref{fig:alpha_L0.1_no_reion}, we show monopoles computed from Landy Szalay estimators and best fit monopoles ($\xi (\alpha^{\ast})$) for the aforementioned mock cuboids. These cuboids have the following minimum values of $\chi^2$: 19.49, 18.58, 11.79, 13.34. They have the following values of $\alpha^{\ast}$: 1.01, 1.03, 0.99, 0.95. Consequently the mean values of minimum $\chi^2$ and $\alpha^{\ast}$ for these mock cuboids are 15.80 and 1.00 respectively.
The fractional distance error is 0.015.

\begin{figure*}
    \includegraphics[width=0.48\textwidth]{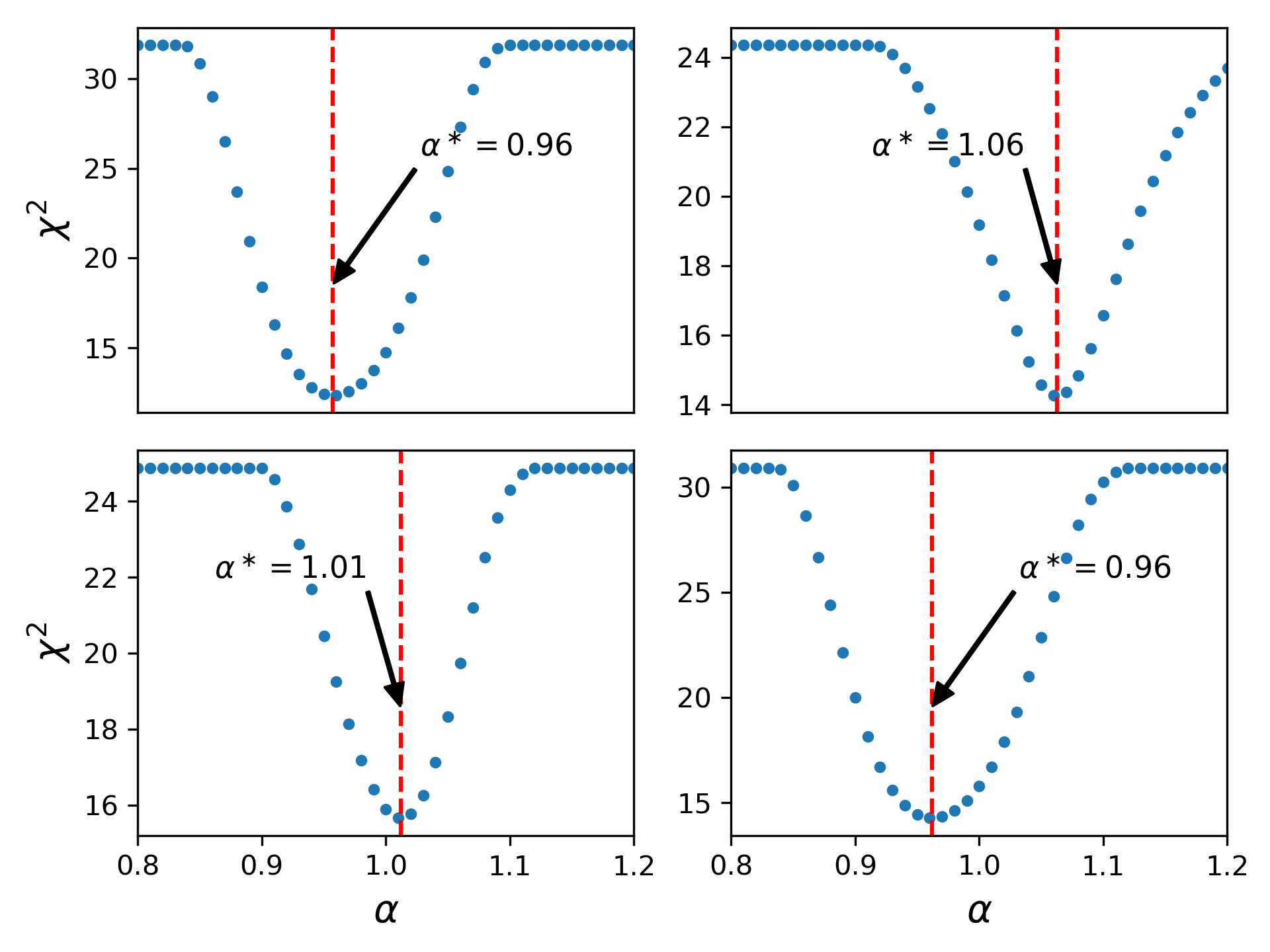}
    \includegraphics[width=0.48\textwidth]{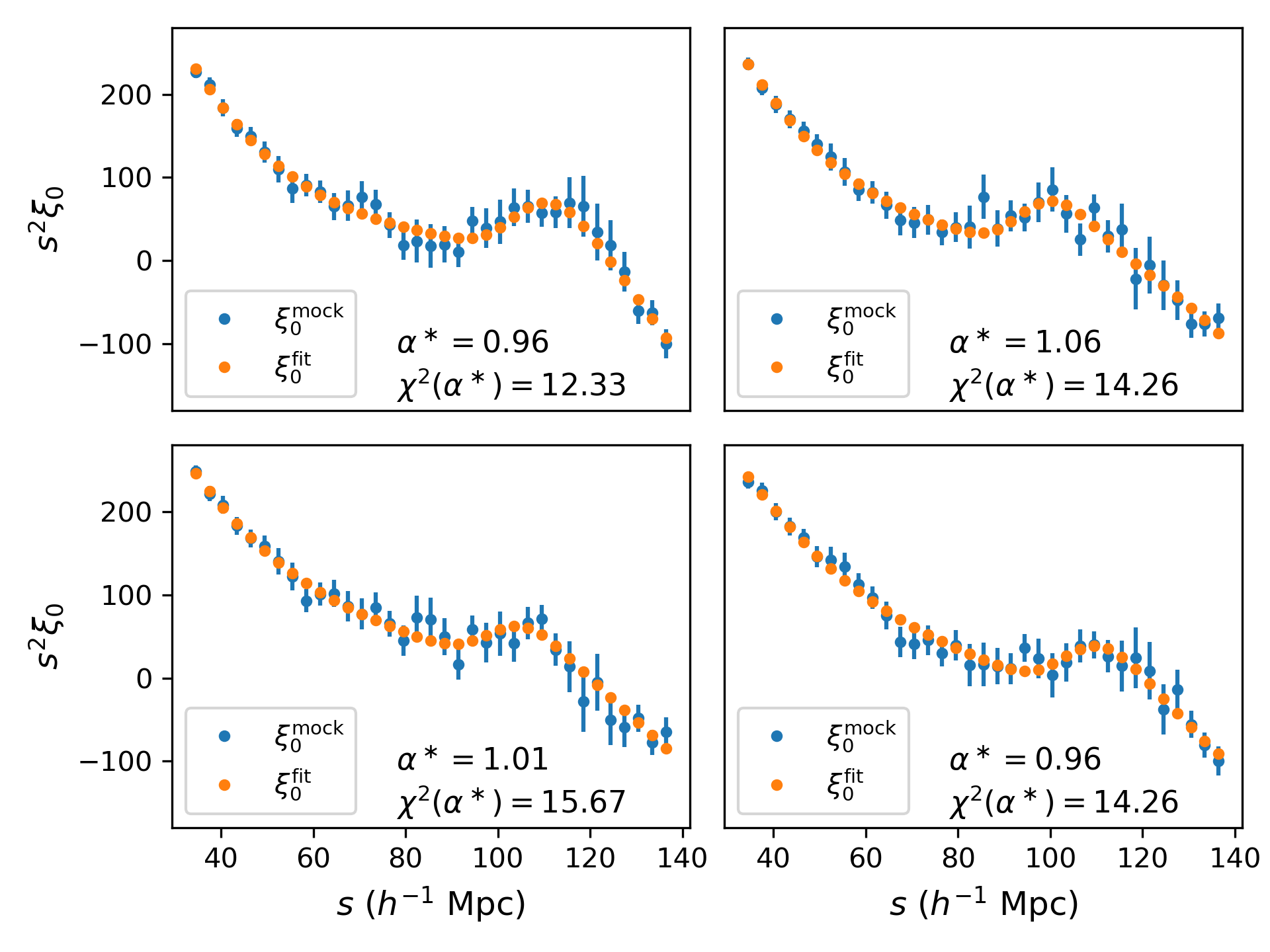}
    \caption{In the left four panels of this figure we show the variations of $\chi^2$ with $\alpha$ for the four mock cuboids corresponding to $L = 1\times 10^{-17} \ {\rm erg/s/cm}^{2}$ (with patchy reionization). In the plots on the left, the red dashed vertical lines denote the positions of the best fit $\alpha$, \textit{viz.} $\alpha^{\ast}$ for different mocks. In the right four panels, we show the monopoles computed from Landy Szalay estimator ($\xi_0^{\rm mock}$) and the best fit monopoles ($\xi_0^{\rm fit}(\alpha^{\ast})$) obtained after fitting of $\xi_0^{\rm mock}$ for the four mock cuboids corresponding to $L = 1\times 10^{-17} \ {\rm erg/s/cm}^{2}$ (with patchy reionization). In the plots in the right panels, the blue dots represent $\xi_0^{\rm mock}$ while the orange dots denote $\xi_0^{\rm fit}(\alpha^{\ast})$.}
    \label{fig:alpha_L0.1_post_reion}
\end{figure*}

The left four panels of Fig.~\ref{fig:alpha_L0.1_post_reion} illustrate the variations of $\chi^2$ with $\alpha$ for the four mock cuboids in L1 quarter-mock catalogue which correspond to distribution of galaxies visible with patchy reionization. In the right four panels of Fig~\ref{fig:alpha_L0.1_post_reion}, we show monopoles computed from Landy Szalay estimators and best fit monopoles ($\xi (\alpha^{\ast})$) for these four mock cuboids. The cuboids L1 quarter-mock catalogue which correspond to distribution of galaxies after patchy reionization have the following values of minimum $\chi^2$: 12.33, 14.26, 15.67, 14.26. The values of $\alpha^{\ast}$ for these cuboids are as follows: 0.96, 1.06, 1.01, 0.96. Hence, the mean values of minimum $\chi^2$ and $\alpha^{\ast}$ for these mock cuboids are 14.13 and 1.00. The fractional
distance error is 0.024.

Table~\ref{tab:summary_results_chi2}, gives summaries of our best fit results for all mocks in all three quarter-mock catalogues.

\begin{table*}
	\centering
	\caption{This table shows best fit results of scale dilation parameter (i.e. $\alpha^{\ast}$) and $\chi^2 (\alpha^{\ast})$ in the three quarter-mock catalogues (L7, L3.5, L1).}
	 \label{tab:summary_results_chi2}
	\begin{tabular}{lcccc}
		\hline
		\hline
    		\textbf{Line detection} & $\alpha^{\ast}$ & $\alpha^{\ast}$  & $\alpha^{\ast}$  & $\alpha^{\ast}$ \\
    		\textbf{sensitivity} & (\textbf{Mock 1}) & (\textbf{Mock 2}) & (\textbf{Mock 3})  & (\textbf{Mock 4})  \\    
    		\hline
    		\hline
		\textbf{Before patchy reionization} & & & & \\    	
    		\hline			
    		$7$ erg/s/cm$^2$ & 1.08 & 1.16 & 0.86 & 1.08 \\        
    		$3.5$ erg/s/cm$^2$ & 0.99 & 1.08 & 1.00 & 1.12 \\        
    		$1$ erg/s/cm$^2$ & 1.01 & 1.03 & 0.99 & 0.95 \\
    		\hline
    		\textbf{With patchy reionization} & & & & \\
    		\hline
    		$7$ erg/s/cm$^2$ & 1.04 & 0.81 & 1.05 & 1.19 \\        
    		$3.5$ erg/s/cm$^2$ & 0.95 & 0.99 & 1.02 & 0.98 \\        
    		$1$ erg/s/cm$^2$ & 0.96 & 1.06 & 1.01 & 0.96 \\
    		\hline   
    		\hline
    		\textbf{Line detection} & $\chi^2 (\alpha^{\ast})$ & $\chi^2 (\alpha^{\ast})$  & $\chi^2 (\alpha^{\ast})$  & $\chi^2 (\alpha^{\ast})$ \\
    		\textbf{sensitivity} & (\textbf{Mock 1}) & (\textbf{Mock 2}) & (\textbf{Mock 3})  & (\textbf{Mock 4})  \\    
    		\hline
    		\hline
		\textbf{Before patchy reionization} & & & & \\    	
    		\hline			
    		$7$ erg/s/cm$^2$ & 27.41 & 31.32 & 28.17 & 45.86 \\        
    		$3.5$ erg/s/cm$^2$ & 21.59 & 20.03 & 16.20 & 26.42 \\        
    		$1$ erg/s/cm$^2$ & 19.49 & 18.58 & 11.79 & 13.34 \\
    		\hline
    		\textbf{With patchy reionization} & & & & \\
    		\hline
    		$7$ erg/s/cm$^2$ & 24.41 & 26.68 & 34.96 & 26.59 \\        
    		$3.5$ erg/s/cm$^2$ & 38.80 & 22.52 & 43.14 & 32.26 \\        
    		$1$ erg/s/cm$^2$ & 12.33 & 14.26 & 15.67 & 14.26 \\
    		\hline   
    		\hline    		    		 		
	 \end{tabular}
\end{table*}

\begin{table*}
	\centering
	\caption{The relative clustering linear bias $b_{\rm reion}$ between mocks with and without reionization in the different quarter-mock catalogues (L7, L3.5, L1).}
	 \label{tab:linear_bias}
	\begin{tabular}{lrrrr}
		\hline
		\hline
    		\textbf{Line detection} & \textbf{Mock 1} & \textbf{Mock 2}  & \textbf{Mock 3}  & \textbf{Mock 4}  \\
    		\textbf{sensitivity} &  $b_{\rm reion}$& $b_{\rm reion}$& $b_{\rm reion}$&  $b_{\rm reion}$\\ 
    		\hline
    		\hline    		
    		$7$ erg/s/cm$^2$ & 1.71 & 1.10 & 1.30 & 1.17 \\        
    		$3.5$ erg/s/cm$^2$ & 1.13 & 1.38 & 1.41 & 1.05 \\        
    		$1$ erg/s/cm$^2$ & 3.76 & 1.52 & 2.65 & 1.70 \\
    		\hline
    		\hline    		
	 \end{tabular}
\end{table*}

\subsection{Analysis of errors on the BAO distance measurements}\label{sec: fisher_test}
We now carry out an analysis of the BAO distance errors and their scatter, in order to check that they are statistically consistent. 
For the L7 quarter mock catalogues, we found the following values of best fit $\alpha$ ($\alpha^{\ast}$) from the four cuboids that represent the distribution of halos without patchy reionization: 1.08, 1.16, 0.86, 1.08. These values of $\alpha^{\ast}$ lead to an overall fractional error of 0.064. At the same time, we observed the following values of $\alpha^{\ast}$ in the four cuboids in the L7 quarter mock catalogues with patchy reionization : 1.04, 0.81, 1.05, 1.19. These values of $\alpha^{\ast}$ yield an error of 0.079 on BAO distance measurement. To reconcile the scatter in the values of $\alpha^{\ast}$ that we see for cuboids with and without patchy reionization, we carry out Fisher's exact test \citep[][]{Fisher1922, Agresti1992}. For each category of cuboids (with reionization or without), we are working with only four cuboids. This is a classic case of data with small sample sizes. Fisher's exact test enables us to conduct accurate statistical significance tests on small samples. In this test, our null hypothesis is that there is no significant deviation in the distributions of the values of $\alpha^{\ast}$ for cases with and without patchy reionization. A two-tailed Fisher exact hypothesis test for comparison of the distributions of the aforesaid values, gives us a p-value which is approximately 0.99. This p value is very high and it strongly suggests that we cannot ignore the null hypothesis.

Similarly, for L3.5 mocks, we have the following values of $\alpha^{\ast}$ for halos without patchy reionization: 0.99, 1.08, 1.00, 1.12. For the four cuboids with patchy reionization, we have: 0.95, 0.99, 1.02, 0.98. We again conduct Fisher's exact test. Our null hypothesis in this test is that there is no significant deviation in distributions of values of $\alpha^{\ast}$. From a two-tailed Fisher's exact test, this time we get a p-value of approximately 1. This strongly suggests that there is significant evidence in support of our null hypothesis.  

Finally, for L1 mocks, we get the following values of $\alpha^{\ast}$ for halos without patchy reionization: 1.01, 1.03, 0.99, 0.95. When cuboids have patchy reionization, we get the following values of $\alpha^{\ast}$: 0.96, 1.06, 1.01, 0.96.  A two-tailed Fisher's exact test gives us a p-value of approximately 1, again strong evidence of the fact that there is no significant deviation in distributions of values of $\alpha^{\ast}$ for cases with and without patchy reionization.

The results that we get from Fisher's exact tests carried out on the mocks also suggest that we cannot make significant inferences from comparisons of scatter in values of $\alpha^{\ast}$ in cuboids with patchy reionization and without patchy reionization.

\subsection{Linear bias between in cases with and without reionization }\label{sec: linear_bias_reion}

The results that we see in Figures \ref{fig:alpha_L0.5_post_reion}, \ref{fig:alpha_L0.5_no_reion}, \ref{fig:alpha_L0.3_post_reion}, \ref{fig:alpha_L0.3_no_reion}, \ref{fig:alpha_L0.1_post_reion} and \ref{fig:alpha_L0.1_no_reion} suggest that the clustering amplitude changes somewhat significantly with and without reionization and also depends on the line strength. Equation~\ref{eqn:linear_bias_reionization} below defines the linear bias due to reionization compared to no reionization for a given cuboid. It is defined using a ratio of two galaxy correlation functions and is of course different from the usual linear bias of galaxies with respect to mass.

\begin{align}
\label{eqn:linear_bias_reionization}
b_{\rm reion} =  \sqrt{ \frac{ \xi_0^{ {\rm reionization} } }{  \xi_0^{ {\rm no \ reionization} } } }
\end{align}

We tabulate the values of $b_{\rm reion}$ for cuboids in the L7, L3.5 and L1 mocks in Table~\ref{tab:linear_bias}. For the L7 mock catalogue, we obtain a mean $b_{\rm reion}$ of 1.32. We estimate the fractional errors on $b_{\rm reion}$ by computing the sample standard deviation of the $b_{\rm reion}$ results for the four quarter-mocks, and then dividing by $\sqrt{4}$ to give the error on the mean. The fractional error computed this way is 0.136. In the case of the L3.5 mock catalogue, we get a mean $b_{\rm reion}$ of 1.24 and a fractional error of 0.089. Finally, for the four quarter-mocks in L1 mock catalogue, we obtain a mean $b_{\rm reion}$ of 2.41 and fractional error of 0.514. It is as we might expect: in all cases the effects of patchy reionization lead to higher amplitude clustering for the observable galaxies. The fainter galaxies (the L1 mocks) are also relatively much more strongly affected than the brighter galaxies.

\section{Summary and Discussion} \label{sec: baodiscussion}

\subsection{Summary}

Using mock catalogues, we have investigated the prospect of utilizing the BAO feature at redshifts $z>7.6$  as a standard ruler. We have utilized four mock catalogues, each representing approximately one quarter of the volume that will be surveyed by RST with grism spectroscopy.

We see evidence for BAO peaks in three of the four mocks when we use the planned RST HLS \lya\ line detection sensitivity of $L = 7\times 10^{-17} \ {\rm erg/s/cm}^{2}$. Hence, the BAO feature is likely detectable for this line detection sensitivity (measured in three out of four survey quadrants analysed independently). When we make mocks with better line detection sensitivities, we see evidence of BAO features in all mocks. For the standard line detection sensitivity, $L = 7\times 10^{-17} \ {\rm erg/s/cm}^{2}$, we  find a root mean square fractional distance error of 0.068 after comparison between mocks and fitted correlation function monopoles. For a factor of two fainter line detection sensitivity, $L = 3.5\times 10^{-17} \ {\rm erg/s/cm}^{2}$, we find a distance error of 0.013. We find a distance error of 0.021 for mocks with line detection sensitivity $L = 1\times 10^{-17} \ {\rm erg/s/cm}^{2}$.
\label{summarysec}

\subsection{Discussion}
Our analysis of the use of BAO at high redshifts has included some of the relevant features and difficulties, such as non-linear clustering, redshift distortions, intrinsic galaxy line luminosities, and the effect of reionization bubbles. The results are promising, as we have summarised in Section~\ref{summarysec} above.  We have however left much for future work, including the visibility of the \lya\ line at these redshifts (including the effects of dust and the \lya\ damping wing), and the effects of interlopers and low redshift foregrounds. These factors could easily derail the possibility  of measuring large-scale galaxy clustering at redshift $z>7.6$, and should be investigated in detail. On the other hand, there are examples of galaxies being detected in \lya\ emission at these redshifts, for example the $z=7.51$ object by \cite{Tilvi2016}, and $z=7.452$ from \cite{Larson2018}. Even if the \lya\ emission line itself cannot be seen, the \lya\ break can also be used to spectroscopically identify galaxies, as was done using the HST grism at $z=11.1$ by \cite{Oesch2016}. This could also be investigated to see whether it could enable clustering measurements at the BAO scale.

We have seen that the detectability of the BAO peak is just at the level of being possible with the fiducial flux limit of the RST HLS. To do this, we have already reduced the nominal 7$\sigma$ detection limit to 5$\sigma$ ($7\times 10^{-17} \ {\rm erg/s/cm}^{2}$). The exact survey design choices made when the RST mission is finalized will therefore be extremely relevant to whether the BAO measurement will be feasible. We have found that with a sensitivity a factor of two better the BAO can be seen clearly in four survey quadrants. This means that it could be advantageous to make use of smaller and deeper areas.


The analysis presented in this work was based on only 4 quarter-mocks for each of the tested line detection sensitivities. In the future, large number of mock catalogues should be generated to allow for the build up of a full covariance matrix for the correlation function.  Because of the small number of mocks here, we were only able to use the diagonal elements of the covariance matrix in model fitting. As a result,  the distance error estimates, which were derived from the scatter between mocks, were necessarily  conservative.  In a future work with large numbers of mocks, the extra information available from the off diagonal covariance matrix elements will mean that the fits to the BAO peaks will be improved and the distance errors will be somewhat reduced. The uncertainty in the distance error will also be  better estimated. In the present case, the error on the error on the mean from only four quarter mocks will be large, and this is why the distance scale error on the L1 mocks increases to 2.4\% from 1.4\% for the L3.5 mocks, in spite of a factor of 3 increase in sensitivity.

In our treatment of reionization, we have assumed that galaxies which lie within ionized bubbles will be visible in \lya\ (consistent with the interpretation of \citet{Bagley2017}). We have compared the effects of including this patchy reionization, or not including it (all galaxies visible), and have found that the effect on BAO peak detection is quite small. For example, for the fiducial L7 sensitivity case, the effect of patchy reionization only increases the distance error from 6.4\% to 7.9 \%. This is likely to be because in the patchy model for reionization employed, overdense regions reionize first, and this is where most of the galaxy clustering signal lies. The reionization bubbles are also on smaller scales,  than the BAO signal, and have a range of sizes, and therefore there is no masking of the BAO peak by bubbles. We note however that we have carried out no \lya\ radiative transfer on the scales of individual galaxies, and that the visibility or not of the line in different environments is likely to be much more complex than our simple bubble model.

One significant difference between the BAO measurements at redshifts $z>7$ and those for $z<1$ is the lack of non-linearities in gravitational clustering. For example we can see that in the right panels of Fig.~\ref{fig:alpha_L0.1_no_reion} there is no obvious smoothing of the BAO peak compared to the linear CDM fit. This is an advantage of working at higher redshifts, as no reconstruction techniques \citep[][]{sherwin2019} are needed to recover the maximum benefit. The high redshift galaxies do have a high bias \citep[][]{bhowmick2019} and so their clustering is detectable where clustering of the density field would not be.

Measuring BAO in a new regime would at the very least be useful as a consistency check on measurements that have been made at low redshifts and in the CMB. It is important to fill in the large gap between the \lya\ forest measurements at $z\sim2.2$ \citep[][]{aubourg2015} and the features seen at $z\sim1100$ by Planck and WMAP \citep[][]{planck2019, bennett2013}. Such measurements might shed some light on the tension between BAO and other  Hubble parameter measurements \citep[][]{cuc2019}, and whether early dark energy may be acting before the recombination epoch \citep[][]{verde2019, poulin2019}.

One of the targets of cosmological analysis with the RST satellite is large scale structure measurements from galaxies at redshifts $z\sim1-3$. This will extend the successful redshift space distortion, BAO, and other techniques to significantly earlier epochs than current large scale structure measurements from ground based surveys. Although much attention has rightly been focussed on this redshift range, we have seen in this paper that it may be possible to make an even greater leap, and use the large scale structure traced out by the very first massive galaxies as a cosmological probe. Although many potential pitfalls remain to be investigated, it is worth bearing this in mind when considering the RST survey strategies and parameters.

\section*{Acknowledgements}
The \textsc{BlueTides} simulation was run on the BlueWaters facility at the National Center for Supercomputing Applications.
TDM acknowledges funding from NSF ACI-1614853, NSF AST-1517593, NSF AST-1616168 and NASA ATP 19-ATP19-0084.
TDM and RACC also acknowledge funding from  NASA ATP 80NSSC18K101, and NASA ATP NNX17AK56G.
RACC was supported by a Lyle Fellowship from the University of Melbourne, and TDM by a Shimmins Fellowship
and a Lyle Fellowship from the University of Melbourne. G.R. acknowledges support from the National Research Foundation of Korea (NRF) through Grant No. 2017077508 funded by the Korean Ministry of Education, Science and Technology (MoEST), and from the faculty research fund of Sejong University in 2018. The work by KH was supported under the U.S. Department of Energy contract DE-AC02-06CH11357. This research used resources of the Argonne Leadership Computing Facility at the Argonne National Laboratory, which is supported by the Office of Science of the U.S. Department of Energy under Contract No. DE-AC02-06CH11357.

\section*{Data availability}
The data underlying this article will be shared on reasonable request to the corresponding author.

\bibliographystyle{mnras}
\bibliography{ref} 

\begin{thebibliography}{}
\makeatletter
\relax
\def\mn@urlcharsother{\let\do\@makeother \do\$\do\&\do\#\do\^\do\_\do\%\do\~}
\def\mn@doi{\begingroup\mn@urlcharsother \@ifnextchar [ {\mn@doi@}
  {\mn@doi@[]}}
\def\mn@doi@[#1]#2{\def\@tempa{#1}\ifx\@tempa\@empty \href
  {http://dx.doi.org/#2} {doi:#2}\else \href {http://dx.doi.org/#2} {#1}\fi
  \endgroup}
\def\mn@eprint#1#2{\mn@eprint@#1:#2::\@nil}
\def\mn@eprint@arXiv#1{\href {http://arxiv.org/abs/#1} {{\tt arXiv:#1}}}
\def\mn@eprint@dblp#1{\href {http://dblp.uni-trier.de/rec/bibtex/#1.xml}
  {dblp:#1}}
\def\mn@eprint@#1:#2:#3:#4\@nil{\def\@tempa {#1}\def\@tempb {#2}\def\@tempc
  {#3}\ifx \@tempc \@empty \let \@tempc \@tempb \let \@tempb \@tempa \fi \ifx
  \@tempb \@empty \def\@tempb {arXiv}\fi \@ifundefined
  {mn@eprint@\@tempb}{\@tempb:\@tempc}{\expandafter \expandafter \csname
  mn@eprint@\@tempb\endcsname \expandafter{\@tempc}}}

\bibitem[\protect\citeauthoryear{Agresti}{Agresti}{1992}]{Agresti1992}
Agresti A.,  1992, \mn@doi [Statist. Sci.] {10.1214/ss/1177011454}, 7, 131

\bibitem[\protect\citeauthoryear{Anderson et~al.,}{Anderson
  et~al.}{2012}]{Anderson2012}
Anderson L.,  et~al., 2012, \mn@doi [Monthly Notices of the Royal Astronomical
  Society] {10.1111/j.1365-2966.2012.22066.x}, 427, 3435

\bibitem[\protect\citeauthoryear{{Ansarinejad} \& {Shanks}}{{Ansarinejad} \&
  {Shanks}}{2018}]{Ansarinejad2018}
{Ansarinejad} B.,  {Shanks} T.,  2018, \mn@doi [Monthly Notices of the Royal
  Astronomical Society] {10.1093/mnras/sty1740}, \href
  {https://ui.adsabs.harvard.edu/abs/2018MNRAS.479.4091A} {479, 4091}

\bibitem[\protect\citeauthoryear{{Ata} et~al.,}{{Ata} et~al.}{2018}]{Ata2018}
{Ata} M.,  et~al., 2018, \mn@doi [\mnras] {10.1093/mnras/stx2630}, \href
  {http://adsabs.harvard.edu/abs/2018MNRAS.473.4773A} {473, 4773}

\bibitem[\protect\citeauthoryear{{Aubourg} et~al.,}{{Aubourg}
  et~al.}{2015}]{aubourg2015}
{Aubourg} {\'E}.,  et~al., 2015, \mn@doi [\prd] {10.1103/PhysRevD.92.123516},
  \href {https://ui.adsabs.harvard.edu/abs/2015PhRvD..92l3516A} {92, 123516}

\bibitem[\protect\citeauthoryear{{Bagley} et~al.,}{{Bagley}
  et~al.}{2017}]{Bagley2017}
{Bagley} M.~B.,  et~al., 2017, \mn@doi [\apj] {10.3847/1538-4357/837/1/11},
  \href {https://ui.adsabs.harvard.edu/abs/2017ApJ...837...11B} {837, 11}

\bibitem[\protect\citeauthoryear{Battaglia, Trac, Cen  \& Loeb}{Battaglia
  et~al.}{2013}]{Battaglia2013}
Battaglia N.,  Trac H.,  Cen R.,   Loeb A.,  2013, \mn@doi [The Astrophysical
  Journal] {10.1088/0004-637x/776/2/81}, 776, 81

\bibitem[\protect\citeauthoryear{{Bautista} et~al.,}{{Bautista}
  et~al.}{2018}]{Bautista2018}
{Bautista} J.~E.,  et~al., 2018, \mn@doi [\apj] {10.3847/1538-4357/aacea5},
  \href {http://adsabs.harvard.edu/abs/2018ApJ...863..110B} {863, 110}

\bibitem[\protect\citeauthoryear{{Bennett} et~al.,}{{Bennett}
  et~al.}{2013}]{bennett2013}
{Bennett} C.~L.,  et~al., 2013, \mn@doi [\apjs] {10.1088/0067-0049/208/2/20},
  \href {https://ui.adsabs.harvard.edu/abs/2013ApJS..208...20B} {208, 20}

\bibitem[\protect\citeauthoryear{Bertin \& Arnouts}{Bertin \&
  Arnouts}{1996}]{Bertin1996}
Bertin E.,  Arnouts S.,  1996, \mn@doi [Astronomy and Astrophysics Supplement
  Series] {10.1051/aas:1996164}, 117, 393

\bibitem[\protect\citeauthoryear{{Bhowmick}, {Somerville}, {DiMatteo},
  {Wilkins}, {Feng}  \& {Tenneti}}{{Bhowmick} et~al.}{2019}]{bhowmick2019}
{Bhowmick} A.~K.,  {Somerville} R.~S.,  {DiMatteo} T.,  {Wilkins} S.,  {Feng}
  Y.,   {Tenneti} A.,  2019, arXiv e-prints, \href
  {https://ui.adsabs.harvard.edu/abs/2019arXiv190802787B} {p. arXiv:1908.02787}

\bibitem[\protect\citeauthoryear{{Bouwens} et~al.,}{{Bouwens}
  et~al.}{2011}]{Bouwens2011}
{Bouwens} R.~J.,  et~al., 2011, \mn@doi [\nat] {10.1038/nature09717}, \href
  {https://ui.adsabs.harvard.edu/abs/2011Natur.469..504B} {469, 504}

\bibitem[\protect\citeauthoryear{{Busca} et~al.,}{{Busca}
  et~al.}{2013}]{Busca2013}
{Busca} N.~G.,  et~al., 2013, \mn@doi [\aap] {10.1051/0004-6361/201220724},
  \href {http://adsabs.harvard.edu/abs/2013A%26A...552A..96B} {552, A96}

\bibitem[\protect\citeauthoryear{Casertano, Brammer, Dixon, Mackenty, Pirzkal,
  Ravindranath  \& Ryan}{Casertano et~al.}{2015}]{Casertano2015}
Casertano S.,  Brammer G.~B.,  Dixon V.,  Mackenty J.~W.,  Pirzkal N.,
  Ravindranath S.,   Ryan R.~E.,  2015, Technical report, {Slitless Grism
  Spectroscopy with WFIRST: Observing Modes and Strategies}, \url
  {http://www.stsci.edu/files/live/sites/www/files/home/scientific-community/wfirst/{\_}documents/WFIRST-STScI-TR1506A.pdf}.
\url
  {http://www.stsci.edu/files/live/sites/www/files/home/scientific-community/wfirst/{\_}documents/WFIRST-STScI-TR1506A.pdf}

\bibitem[\protect\citeauthoryear{{Cassata, P.} et~al.,}{{Cassata, P.}
  et~al.}{2011}]{Cassata2011}
{Cassata, P.} et~al., 2011, \mn@doi [A\&A] {10.1051/0004-6361/201014410}, 525,
  A143

\bibitem[\protect\citeauthoryear{{Chaikin}, {Tyulneva}  \& {Kaurov}}{{Chaikin}
  et~al.}{2018}]{Chaikin2018}
{Chaikin} E.~A.,  {Tyulneva} N.~V.,   {Kaurov} A.~A.,  2018, \mn@doi [\apj]
  {10.3847/1538-4357/aaa196}, \href
  {https://ui.adsabs.harvard.edu/abs/2018ApJ...853...81C} {853, 81}

\bibitem[\protect\citeauthoryear{{Colbert} et~al.,}{{Colbert}
  et~al.}{2018}]{Colbert2018}
{Colbert} J.,  et~al., 2018, in American Astronomical Society Meeting Abstracts
  \#231. p. 355.12

\bibitem[\protect\citeauthoryear{{Cole} et~al.,}{{Cole}
  et~al.}{2005}]{Cole2005}
{Cole} S.,  et~al., 2005, \mn@doi [\mnras] {10.1111/j.1365-2966.2005.09318.x},
  \href {http://adsabs.harvard.edu/abs/2005MNRAS.362..505C} {362, 505}

\bibitem[\protect\citeauthoryear{{Comparat} et~al.,}{{Comparat}
  et~al.}{2016}]{Comparat2016}
{Comparat} J.,  et~al., 2016, \mn@doi [\aap] {10.1051/0004-6361/201527377},
  \href {https://ui.adsabs.harvard.edu/abs/2016A%26A...592A.121C} {592, A121}

\bibitem[\protect\citeauthoryear{Crocce \& Scoccimarro}{Crocce \&
  Scoccimarro}{2008}]{Crocce2008}
Crocce M.,  Scoccimarro R.,  2008, \mn@doi [Phys. Rev. D]
  {10.1103/PhysRevD.77.023533}, 77, 023533

\bibitem[\protect\citeauthoryear{{Cuceu}, {Farr}, {Lemos}  \&
  {Font-Ribera}}{{Cuceu} et~al.}{2019}]{cuc2019}
{Cuceu} A.,  {Farr} J.,  {Lemos} P.,   {Font-Ribera} A.,  2019, \mn@doi [\jcap]
  {10.1088/1475-7516/2019/10/044}, \href
  {https://ui.adsabs.harvard.edu/abs/2019JCAP...10..044C} {2019, 044}

\bibitem[\protect\citeauthoryear{{Davis} \& {Peebles}}{{Davis} \&
  {Peebles}}{1983}]{DavisPeebles1983}
{Davis} M.,  {Peebles} P.~J.~E.,  1983, \mn@doi [\apj] {10.1086/160884}, \href
  {http://adsabs.harvard.edu/abs/1983ApJ...267..465D} {267, 465}

\bibitem[\protect\citeauthoryear{{Davis}, {Efstathiou}, {Frenk}  \&
  {White}}{{Davis} et~al.}{1985}]{Davis1985}
{Davis} M.,  {Efstathiou} G.,  {Frenk} C.~S.,   {White} S.~D.~M.,  1985,
  \mn@doi [\apj] {10.1086/163168}, \href
  {https://ui.adsabs.harvard.edu/abs/1985ApJ...292..371D} {292, 371}

\bibitem[\protect\citeauthoryear{{Eftekharzadeh} et~al.,}{{Eftekharzadeh}
  et~al.}{2015}]{Eftekharzadeh2015}
{Eftekharzadeh} S.,  et~al., 2015, \mn@doi [Monthly Notices of the Royal
  Astronomical Society] {10.1093/mnras/stv1763}, \href
  {https://ui.adsabs.harvard.edu/abs/2015MNRAS.453.2779E} {453, 2779}

\bibitem[\protect\citeauthoryear{{Eisenstein} \& {Hu}}{{Eisenstein} \&
  {Hu}}{1998}]{Eisenstein1998}
{Eisenstein} D.~J.,  {Hu} W.,  1998, \mn@doi [\apj] {10.1086/305424}, \href
  {https://ui.adsabs.harvard.edu/abs/1998ApJ...496..605E} {496, 605}

\bibitem[\protect\citeauthoryear{{Eisenstein} et~al.,}{{Eisenstein}
  et~al.}{2005}]{Eisenstein2005}
{Eisenstein} D.~J.,  et~al., 2005, \mn@doi [\apj] {10.1086/466512}, \href
  {http://adsabs.harvard.edu/abs/2005ApJ...633..560E} {633, 560}

\bibitem[\protect\citeauthoryear{{Ellis} et~al.,}{{Ellis}
  et~al.}{2013}]{Ellis2013}
{Ellis} R.~S.,  et~al., 2013, \mn@doi [\apjl] {10.1088/2041-8205/763/1/L7},
  \href {https://ui.adsabs.harvard.edu/abs/2013ApJ...763L...7E} {763, L7}

\bibitem[\protect\citeauthoryear{{Eyles}, {Bunker}, {Stanway}, {Lacy}, {Ellis}
  \& {Doherty}}{{Eyles} et~al.}{2005}]{Eyles2005}
{Eyles} L.~P.,  {Bunker} A.~J.,  {Stanway} E.~R.,  {Lacy} M.,  {Ellis} R.~S.,
  {Doherty} M.,  2005, \mn@doi [\mnras] {10.1111/j.1365-2966.2005.09434.x},
  \href {https://ui.adsabs.harvard.edu/abs/2005MNRAS.364..443E} {364, 443}

\bibitem[\protect\citeauthoryear{Feng, Matteo, Croft, Tenneti, Bird, Battaglia
  \& Wilkins}{Feng et~al.}{2015}]{Feng2015}
Feng Y.,  Matteo T.~D.,  Croft R.,  Tenneti A.,  Bird S.,  Battaglia N.,
  Wilkins S.,  2015, \mn@doi [The Astrophysical Journal]
  {10.1088/2041-8205/808/1/l17}, 808, L17

\bibitem[\protect\citeauthoryear{Feng, Di-Matteo, Croft, Bird, Battaglia  \&
  Wilkins}{Feng et~al.}{2016}]{Feng2016}
Feng Y.,  Di-Matteo T.,  Croft R.~A.,  Bird S.,  Battaglia N.,   Wilkins S.,
  2016, \mn@doi [Monthly Notices of the Royal Astronomical Society]
  {10.1093/mnras/stv2484}, 455, 2778

\bibitem[\protect\citeauthoryear{{Finkelstein} et~al.,}{{Finkelstein}
  et~al.}{2013}]{Finkelstein2013}
{Finkelstein} S.~L.,  et~al., 2013, \mn@doi [\nat] {10.1038/nature12657}, \href
  {https://ui.adsabs.harvard.edu/abs/2013Natur.502..524F} {502, 524}

\bibitem[\protect\citeauthoryear{Fisher}{Fisher}{1922}]{Fisher1922}
Fisher R.~A.,  1922, Journal of the Royal Statistical Society, 85, 87

\bibitem[\protect\citeauthoryear{{Fossati} et~al.,}{{Fossati}
  et~al.}{2017}]{Fossati2017}
{Fossati} M.,  et~al., 2017, \mn@doi [The Astrophysical Journal]
  {10.3847/1538-4357/835/2/153}, \href
  {https://ui.adsabs.harvard.edu/abs/2017ApJ...835..153F} {835, 153}

\bibitem[\protect\citeauthoryear{Francis, Dopita, Colbert, Palunas, Scarlata,
  Teplitz, Williger  \& Woodgate}{Francis et~al.}{2012}]{Francis2012}
Francis P.~J.,  Dopita M.~A.,  Colbert J.~W.,  Palunas P.,  Scarlata C.,
  Teplitz H.,  Williger G.~M.,   Woodgate B.~E.,  2012, \mn@doi [Monthly
  Notices of the Royal Astronomical Society] {10.1093/mnras/sts010}, 428, 28

\bibitem[\protect\citeauthoryear{{Gehrels}, {Spergel}  \& {WFIRST SDT
  Project}}{{Gehrels} et~al.}{2015}]{Gehrels2015}
{Gehrels} N.,  {Spergel} D.,   {WFIRST SDT Project} 2015, in Journal of Physics
  Conference Series. p. 012007 (\mn@eprint {arXiv} {1411.0313}),
  \mn@doi{10.1088/1742-6596/610/1/012007}

\bibitem[\protect\citeauthoryear{Gronwall et~al.,}{Gronwall
  et~al.}{2007}]{Gronwall_2007a}
Gronwall C.,  et~al., 2007, \mn@doi [The Astrophysical Journal]
  {10.1086/520324}, 667, 79

\bibitem[\protect\citeauthoryear{{Habib}, {Morozov}, {Frontiere}, {Finkel},
  {Pope}  \& {Heitmann}}{{Habib} et~al.}{2013}]{Habib2013}
{Habib} S.,  {Morozov} V.,  {Frontiere} N.,  {Finkel} H.,  {Pope} A.,
  {Heitmann} K.,  2013, in SC '13: Proceedings of the International Conference
  on High Performance Computing, Networking, Storage and Analysis. pp 1--10,
  \mn@doi{10.1145/2503210.2504566}

\bibitem[\protect\citeauthoryear{{Habib} et~al.,}{{Habib}
  et~al.}{2016}]{Habib2016}
{Habib} S.,  et~al., 2016, \mn@doi [New Astronomy]
  {10.1016/j.newast.2015.06.003}, \href
  {https://ui.adsabs.harvard.edu/abs/2016NewA...42...49H} {42, 49}

\bibitem[\protect\citeauthoryear{{Hamilton}}{{Hamilton}}{1993}]{Hamilton1993}
{Hamilton} A.~J.~S.,  1993, \mn@doi [\apj] {10.1086/173288}, \href
  {http://adsabs.harvard.edu/abs/1993ApJ...417...19H} {417, 19}

\bibitem[\protect\citeauthoryear{{Heitmann} et~al.,}{{Heitmann}
  et~al.}{2019}]{Heitmann2019}
{Heitmann} K.,  et~al., 2019, arXiv e-prints, \href
  {https://ui.adsabs.harvard.edu/abs/2019arXiv190411970H} {p. arXiv:1904.11970}

\bibitem[\protect\citeauthoryear{{Hill} \& {Baxter}}{{Hill} \&
  {Baxter}}{2018}]{Hill2018}
{Hill} J.~C.,  {Baxter} E.~J.,  2018, \mn@doi [\jcap]
  {10.1088/1475-7516/2018/08/037}, \href
  {http://adsabs.harvard.edu/abs/2018JCAP...08..037H} {8, 037}

\bibitem[\protect\citeauthoryear{{Hu} et~al.,}{{Hu} et~al.}{2017}]{Hu2017}
{Hu} W.,  et~al., 2017, \mn@doi [\apjl] {10.3847/2041-8213/aa8401}, \href
  {https://ui.adsabs.harvard.edu/abs/2017ApJ...845L..16H} {845, L16}

\bibitem[\protect\citeauthoryear{{Kakiichi}, {Dijkstra}, {Ciardi}  \&
  {Graziani}}{{Kakiichi} et~al.}{2016}]{Kakiichi2016}
{Kakiichi} K.,  {Dijkstra} M.,  {Ciardi} B.,   {Graziani} L.,  2016, \mn@doi
  [Monthly Notices of the Royal Astronomical Society] {10.1093/mnras/stw2193},
  \href {https://ui.adsabs.harvard.edu/abs/2016MNRAS.463.4019K} {463, 4019}

\bibitem[\protect\citeauthoryear{Kerscher, Szapudi  \& Szalay}{Kerscher
  et~al.}{2000}]{Kerscher2000}
Kerscher M.,  Szapudi I.,   Szalay A.~S.,  2000, \mn@doi [The Astrophysical
  Journal Letters] {10.1086/312702}, \href
  {http://adsabs.harvard.edu/abs/2000ApJ...535L..13K} {535, L13}

\bibitem[\protect\citeauthoryear{Komatsu et~al.,}{Komatsu
  et~al.}{2011}]{Komatsu2011}
Komatsu E.,  et~al., 2011, \mn@doi [Astrophysical Journal, Supplement Series]
  {10.1088/0067-0049/192/2/18}, 192, 18

\bibitem[\protect\citeauthoryear{{Landy} \& {Szalay}}{{Landy} \&
  {Szalay}}{1993}]{Landy1993}
{Landy} S.~D.,  {Szalay} A.~S.,  1993, \mn@doi [\apj] {10.1086/172900}, \href
  {http://adsabs.harvard.edu/abs/1993ApJ...412...64L} {412, 64}

\bibitem[\protect\citeauthoryear{{Laporte} et~al.,}{{Laporte}
  et~al.}{2017}]{Laporte2017}
{Laporte} N.,  et~al., 2017, \mn@doi [\apjl] {10.3847/2041-8213/aa62aa}, \href
  {https://ui.adsabs.harvard.edu/abs/2017ApJ...837L..21L} {837, L21}

\bibitem[\protect\citeauthoryear{{Larson} et~al.,}{{Larson}
  et~al.}{2018}]{Larson2018}
{Larson} R.~L.,  et~al., 2018, \mn@doi [\apj] {10.3847/1538-4357/aab893}, \href
  {https://ui.adsabs.harvard.edu/abs/2018ApJ...858...94L} {858, 94}

\bibitem[\protect\citeauthoryear{{Lehnert} et~al.,}{{Lehnert}
  et~al.}{2010}]{Lehnert2010}
{Lehnert} M.~D.,  et~al., 2010, \mn@doi [Nature] {10.1038/nature09462}, \href
  {https://ui.adsabs.harvard.edu/abs/2010Natur.467..940L} {467, 940}

\bibitem[\protect\citeauthoryear{{Leitherer} et~al.,}{{Leitherer}
  et~al.}{1999}]{Leitherer1999}
{Leitherer} C.,  et~al., 1999, \mn@doi [\apjs] {10.1086/313233}, \href
  {https://ui.adsabs.harvard.edu/abs/1999ApJS..123....3L} {123, 3}

\bibitem[\protect\citeauthoryear{Lewis \& Bridle}{Lewis \&
  Bridle}{2002}]{Lewis2002}
Lewis A.,  Bridle S.,  2002, \mn@doi [Phys. Rev. D]
  {10.1103/PhysRevD.66.103511}, 66, 103511

\bibitem[\protect\citeauthoryear{{Li} et~al.,}{{Li} et~al.}{2019}]{Li2019}
{Li} N.,  et~al., 2019, \mn@doi [\apj] {10.3847/1538-4357/ab1f74}, \href
  {https://ui.adsabs.harvard.edu/abs/2019ApJ...878..122L} {878, 122}

\bibitem[\protect\citeauthoryear{{Malhotra}}{{Malhotra}}{2019}]{Malhotra2019}
{Malhotra} S.,  2019, in American Astronomical Society Meeting Abstracts. p.
  315.01

\bibitem[\protect\citeauthoryear{{McCarthy} et~al.,}{{McCarthy}
  et~al.}{1999}]{McCarthy1999}
{McCarthy} P.~J.,  et~al., 1999, \mn@doi [The Astrophysical Journal]
  {10.1086/307491}, \href
  {https://ui.adsabs.harvard.edu/abs/1999ApJ...520..548M} {520, 548}

\bibitem[\protect\citeauthoryear{{Mobasher} et~al.,}{{Mobasher}
  et~al.}{2005}]{Mobasher2005}
{Mobasher} B.,  et~al., 2005, \mn@doi [\apj] {10.1086/497626}, \href
  {https://ui.adsabs.harvard.edu/abs/2005ApJ...635..832M} {635, 832}

\bibitem[\protect\citeauthoryear{{Oesch} et~al.,}{{Oesch}
  et~al.}{2016}]{Oesch2016}
{Oesch} P.~A.,  et~al., 2016, \mn@doi [\apj] {10.3847/0004-637X/819/2/129},
  \href {http://adsabs.harvard.edu/abs/2016ApJ...819..129O} {819, 129}

\bibitem[\protect\citeauthoryear{{Pentericci} et~al.,}{{Pentericci}
  et~al.}{2018}]{Pentericci2018}
{Pentericci} L.,  et~al., 2018, \mn@doi [\aap] {10.1051/0004-6361/201732465},
  \href {https://ui.adsabs.harvard.edu/abs/2018A%26A...619A.147P} {619, A147}

\bibitem[\protect\citeauthoryear{{Planck Collaboration} et~al.,}{{Planck
  Collaboration} et~al.}{2018a}]{Planck2018}
{Planck Collaboration} et~al., 2018a, arXiv e-prints, \href
  {http://adsabs.harvard.edu/abs/2018arXiv180706209P} {}

\bibitem[\protect\citeauthoryear{{Planck Collaboration} et~al.,}{{Planck
  Collaboration} et~al.}{2018b}]{planck2019}
{Planck Collaboration} et~al., 2018b, arXiv e-prints, \href
  {https://ui.adsabs.harvard.edu/abs/2018arXiv180706209P} {p. arXiv:1807.06209}

\bibitem[\protect\citeauthoryear{Pons-Border\'{i}a, Mart\'{i}nez, Stoyan,
  Stoyan  \& Saar}{Pons-Border\'{i}a et~al.}{1999}]{Pons1999}
Pons-Border\'{i}a M.-J.,  Mart\'{i}nez V.~J.,  Stoyan D.,  Stoyan H.,   Saar
  E.,  1999, \mn@doi [\apj] {10.1086/307754}, 523, 480

\bibitem[\protect\citeauthoryear{{Poulin}, {Smith}, {Karwal}  \&
  {Kamionkowski}}{{Poulin} et~al.}{2019}]{poulin2019}
{Poulin} V.,  {Smith} T.~L.,  {Karwal} T.,   {Kamionkowski} M.,  2019, \mn@doi
  [\prl] {10.1103/PhysRevLett.122.221301}, \href
  {https://ui.adsabs.harvard.edu/abs/2019PhRvL.122v1301P} {122, 221301}

\bibitem[\protect\citeauthoryear{{Raichoor} et~al.,}{{Raichoor}
  et~al.}{2017}]{Raichoor2017}
{Raichoor} A.,  et~al., 2017, \mn@doi [\mnras] {10.1093/mnras/stx1790}, \href
  {http://adsabs.harvard.edu/abs/2017MNRAS.471.3955R} {471, 3955}

\bibitem[\protect\citeauthoryear{{Riess} et~al.,}{{Riess}
  et~al.}{2018}]{Riess2018}
{Riess} A.~G.,  et~al., 2018, \mn@doi [\apj] {10.3847/1538-4357/aaadb7}, \href
  {http://adsabs.harvard.edu/abs/2018ApJ...855..136R} {855, 136}

\bibitem[\protect\citeauthoryear{{Sarpa}, {Schimd}, {Branchini}  \&
  {Matarrese}}{{Sarpa} et~al.}{2019}]{Sarpa2019}
{Sarpa} E.,  {Schimd} C.,  {Branchini} E.,   {Matarrese} S.,  2019, \mn@doi
  [\mnras] {10.1093/mnras/stz278}, \href
  {https://ui.adsabs.harvard.edu/abs/2019MNRAS.484.3818S} {484, 3818}

\bibitem[\protect\citeauthoryear{{Sherwin} \& {White}}{{Sherwin} \&
  {White}}{2019}]{sherwin2019}
{Sherwin} B.~D.,  {White} M.,  2019, \mn@doi [\jcap]
  {10.1088/1475-7516/2019/02/027}, \href
  {https://ui.adsabs.harvard.edu/abs/2019JCAP...02..027S} {2019, 027}

\bibitem[\protect\citeauthoryear{{Slosar} et~al.,}{{Slosar}
  et~al.}{2013}]{Slosar2013}
{Slosar} A.,  et~al., 2013, \mn@doi [\jcap] {10.1088/1475-7516/2013/04/026},
  \href {http://adsabs.harvard.edu/abs/2013JCAP...04..026S} {4, 026}

\bibitem[\protect\citeauthoryear{{Sobacchi} \& {Mesinger}}{{Sobacchi} \&
  {Mesinger}}{2015}]{Sobacchi2015}
{Sobacchi} E.,  {Mesinger} A.,  2015, \mn@doi [Monthly Notices of the Royal
  Astronomical Society] {10.1093/mnras/stv1751}, \href
  {https://ui.adsabs.harvard.edu/abs/2015MNRAS.453.1843S} {453, 1843}

\bibitem[\protect\citeauthoryear{Song, Finkelstein, Livermore, Capak, Dickinson
   \& Fontana}{Song et~al.}{2016}]{Song2016}
Song M.,  Finkelstein S.~L.,  Livermore R.~C.,  Capak P.~L.,  Dickinson M.,
  Fontana A.,  2016, \mn@doi [The Astrophysical Journal]
  {10.3847/0004-637x/826/2/113}, 826, 113

\bibitem[\protect\citeauthoryear{{Spergel} et~al.,}{{Spergel}
  et~al.}{2015}]{spergel2015}
{Spergel} D.,  et~al., 2015, arXiv e-prints, \href
  {https://ui.adsabs.harvard.edu/abs/2015arXiv150303757S} {p. arXiv:1503.03757}

\bibitem[\protect\citeauthoryear{{Tilvi} et~al.,}{{Tilvi}
  et~al.}{2010}]{Tilvi2010}
{Tilvi} V.,  et~al., 2010, \mn@doi [The Astrophysical Journal]
  {10.1088/0004-637X/721/2/1853}, \href
  {https://ui.adsabs.harvard.edu/abs/2010ApJ...721.1853T} {721, 1853}

\bibitem[\protect\citeauthoryear{{Tilvi} et~al.,}{{Tilvi}
  et~al.}{2016}]{Tilvi2016}
{Tilvi} V.,  et~al., 2016, \mn@doi [\apjl] {10.3847/2041-8205/827/1/L14}, \href
  {https://ui.adsabs.harvard.edu/abs/2016ApJ...827L..14T} {827, L14}

\bibitem[\protect\citeauthoryear{{Vanderlinde} \& {Chime
  Collaboration}}{{Vanderlinde} \& {Chime
  Collaboration}}{2014}]{Vanderlinde2014}
{Vanderlinde} K.,  {Chime Collaboration} 2014, in Exascale Radio Astronomy.

\bibitem[\protect\citeauthoryear{Vargas-Maga{\~{n}}a
  et~al.,}{Vargas-Maga{\~{n}}a et~al.}{2014}]{Vargas2014}
Vargas-Maga{\~{n}}a M.,  et~al., 2014, \mn@doi [Monthly Notices of the Royal
  Astronomical Society] {10.1093/mnras/stu1681}, 445, 2

\bibitem[\protect\citeauthoryear{{Verde}, {Treu}  \& {Riess}}{{Verde}
  et~al.}{2019}]{verde2019}
{Verde} L.,  {Treu} T.,   {Riess} A.~G.,  2019, \mn@doi [Nature Astronomy]
  {10.1038/s41550-019-0902-0}, \href
  {https://ui.adsabs.harvard.edu/abs/2019NatAs...3..891V} {3, 891}

\bibitem[\protect\citeauthoryear{{Wang}, {Peter}, {Strigari}, {Zentner},
  {Arant}, {Garrison-Kimmel}  \& {Rocha}}{{Wang} et~al.}{2014}]{Wang2014}
{Wang} M.-Y.,  {Peter} A.~H.~G.,  {Strigari} L.~E.,  {Zentner} A.~R.,  {Arant}
  B.,  {Garrison-Kimmel} S.,   {Rocha} M.,  2014, \mn@doi [\mnras]
  {10.1093/mnras/stu1747}, \href
  {http://adsabs.harvard.edu/abs/2014MNRAS.445..614W} {445, 614}

\bibitem[\protect\citeauthoryear{{Waters}, {Di Matteo}, {Feng}, {Wilkins}  \&
  {Croft}}{{Waters} et~al.}{2016}]{waters2016}
{Waters} D.,  {Di Matteo} T.,  {Feng} Y.,  {Wilkins} S.~M.,   {Croft} R. A.~C.,
   2016, \mn@doi [\mnras] {10.1093/mnras/stw2000}, \href
  {https://ui.adsabs.harvard.edu/abs/2016MNRAS.463.3520W} {463, 3520}

\bibitem[\protect\citeauthoryear{{Weinberg}, {Mortonson}, {Eisenstein},
  {Hirata}, {Riess}  \& {Rozo}}{{Weinberg} et~al.}{2013}]{Weinberg2013}
{Weinberg} D.~H.,  {Mortonson} M.~J.,  {Eisenstein} D.~J.,  {Hirata} C.,
  {Riess} A.~G.,   {Rozo} E.,  2013, \mn@doi [\physrep]
  {10.1016/j.physrep.2013.05.001}, \href
  {http://adsabs.harvard.edu/abs/2013PhR...530...87W} {530, 87}

\bibitem[\protect\citeauthoryear{Wilkins et~al.,}{Wilkins
  et~al.}{2013}]{Wilkins2013b}
Wilkins S.~M.,  et~al., 2013, \mn@doi [Monthly Notices of the Royal
  Astronomical Society] {10.1093/mnras/stt1471}, 435, 2885

\bibitem[\protect\citeauthoryear{{Wilkins}, {Feng}, {Di-Matteo}, {Croft},
  {Stanway}, {Bouwens}  \& {Thomas}}{{Wilkins} et~al.}{2016a}]{Wilkins2016a}
{Wilkins} S.~M.,  {Feng} Y.,  {Di-Matteo} T.,  {Croft} R.,  {Stanway} E.~R.,
  {Bouwens} R.~J.,   {Thomas} P.,  2016a, \mn@doi [\mnras]
  {10.1093/mnrasl/slw007}, \href
  {https://ui.adsabs.harvard.edu/abs/2016MNRAS.458L...6W} {458, L6}

\bibitem[\protect\citeauthoryear{{Wilkins}, {Feng}, {Di-Matteo}, {Croft},
  {Stanway}, {Bunker}, {Waters}  \& {Lovell}}{{Wilkins}
  et~al.}{2016b}]{Wilkins2016b}
{Wilkins} S.~M.,  {Feng} Y.,  {Di-Matteo} T.,  {Croft} R.,  {Stanway} E.~R.,
  {Bunker} A.,  {Waters} D.,   {Lovell} C.,  2016b, \mn@doi [\mnras]
  {10.1093/mnras/stw1154}, \href
  {https://ui.adsabs.harvard.edu/abs/2016MNRAS.460.3170W} {460, 3170}

\bibitem[\protect\citeauthoryear{{Wilkins}, {Feng}, {Di Matteo}, {Croft},
  {Lovell}  \& {Waters}}{{Wilkins} et~al.}{2017}]{Wilkins2017}
{Wilkins} S.~M.,  {Feng} Y.,  {Di Matteo} T.,  {Croft} R.,  {Lovell} C.~C.,
  {Waters} D.,  2017, \mn@doi [\mnras] {10.1093/mnras/stx841}, \href
  {https://ui.adsabs.harvard.edu/abs/2017MNRAS.469.2517W} {469, 2517}

\bibitem[\protect\citeauthoryear{{Wilkins} et~al.,}{{Wilkins}
  et~al.}{2019}]{Wilkins2019}
{Wilkins} S.~M.,  et~al., 2019, arXiv e-prints, \href
  {https://ui.adsabs.harvard.edu/abs/2019arXiv190407504W} {p. arXiv:1904.07504}

\bibitem[\protect\citeauthoryear{{Xu}, {Padmanabhan}, {Eisenstein}, {Mehta}  \&
  {Cuesta}}{{Xu} et~al.}{2012}]{Xu2012}
{Xu} X.,  {Padmanabhan} N.,  {Eisenstein} D.~J.,  {Mehta} K.~T.,   {Cuesta}
  A.~J.,  2012, \mn@doi [\mnras] {10.1111/j.1365-2966.2012.21573.x}, \href
  {http://adsabs.harvard.edu/abs/2012MNRAS.427.2146X} {427, 2146}

\bibitem[\protect\citeauthoryear{{Xu}, {Cuesta}, {Padmanabhan}, {Eisenstein}
  \& {McBride}}{{Xu} et~al.}{2013}]{Xu2013}
{Xu} X.,  {Cuesta} A.~J.,  {Padmanabhan} N.,  {Eisenstein} D.~J.,   {McBride}
  C.~K.,  2013, \mn@doi [\mnras] {10.1093/mnras/stt379}, \href
  {http://adsabs.harvard.edu/abs/2013MNRAS.431.2834X} {431, 2834}

\bibitem[\protect\citeauthoryear{{Yajima}, {Sugimura}  \& {Hasegawa}}{{Yajima}
  et~al.}{2018}]{Yajima2018}
{Yajima} H.,  {Sugimura} K.,   {Hasegawa} K.,  2018, \mn@doi [Monthly Notices
  of the Royal Astronomical Society] {10.1093/mnras/sty997}, \href
  {https://ui.adsabs.harvard.edu/abs/2018MNRAS.477.5406Y} {477, 5406}

\bibitem[\protect\citeauthoryear{{Yan} et~al.,}{{Yan} et~al.}{2011}]{Yan2011}
{Yan} H.,  et~al., 2011, \mn@doi [\apjl] {10.1088/2041-8205/728/1/L22}, \href
  {https://ui.adsabs.harvard.edu/abs/2011ApJ...728L..22Y} {728, L22}

\bibitem[\protect\citeauthoryear{Zarrouk et~al.,}{Zarrouk
  et~al.}{2018}]{Zarrouk2018}
Zarrouk P.,  et~al., 2018, \mn@doi [Monthly Notices of the Royal Astronomical
  Society] {10.1093/mnras/sty506}, 477, 1639

\bibitem[\protect\citeauthoryear{{Zhai} et~al.,}{{Zhai}
  et~al.}{2017}]{Zhai2017}
{Zhai} Z.,  et~al., 2017, \mn@doi [\apj] {10.3847/1538-4357/aa8eee}, \href
  {http://adsabs.harvard.edu/abs/2017ApJ...848...76Z} {848, 76}

\bibitem[\protect\citeauthoryear{{de Sainte Agathe} et~al.,}{{de Sainte Agathe}
  et~al.}{2019}]{deSainte2019}
{de Sainte Agathe} V.,  et~al., 2019, \mn@doi [\aap]
  {10.1051/0004-6361/201935638}, \href
  {https://ui.adsabs.harvard.edu/abs/2019A&A...629A..85D} {629, A85}

\bibitem[\protect\citeauthoryear{{du Mas des Bourboux} et~al.,}{{du Mas des
  Bourboux} et~al.}{2017}]{duMas2017}
{du Mas des Bourboux} H.,  et~al., 2017, \mn@doi [\aap]
  {10.1051/0004-6361/201731731}, \href
  {https://ui.adsabs.harvard.edu/abs/2017A&A...608A.130D} {608, A130}

\makeatother
\end{thebibliography}



\appendix



\bsp	
\label{lastpage}
\end{document}